\documentclass[runningheads,a4paper]{llncs}

\usepackage{amssymb}
\usepackage{graphicx}

\usepackage{url}
\urldef{\mailsa}\path|{alfred.hofmann, ursula.barth, ingrid.haas, frank.holzwarth,|
\urldef{\mailsb}\path|anna.kramer, leonie.kunz, christine.reiss, nicole.sator,|
\urldef{\mailsc}\path|erika.siebert-cole, peter.strasser, lncs}@springer.com|    
\newcommand{\keywords}[1]{\par\addvspace\baselineskip
\noindent\keywordname\enspace\ignorespaces#1}

\begin{document}

\mainmatter  

\title{ Chiral Magnetic   Effect  in
Hydrodynamic Approximation }

\titlerunning{Chiral Magnetic Effect in Hydrodynamics}

%
%
\author{Valentin I. Zakharov%
}
\authorrunning{V.I. Zakharov} 

\institute{Institute of Theoretical and  Experimental Physics\\
B. Cheremushkinskaya, 25, Moscow, Russia and\\
Max-Planck Institut fuer Physik, Werner-Heisenberg Institut\\
Foehringer Ring, 6, Muenchen, Germany\\ 
vzakharov@itep.ru\\
 }

%
%

\toctitle{Lecture Notes in Physics}
\maketitle

\begin{abstract}
We review derivations of the chiral magnetic effect (ChME) in hydrodynamic approximation. 
The reader is assumed to be familiar with the basics of the effect. The main challenge  
now is to account for the strong interactions between the constituents of
the fluid. 
The main result is that the ChME is not renormalized: in the hydrodynamic approximation it remains
the same as for non-interacting chiral fermions moving in an external magnetic field.
The key ingredients in the proof are general laws of thermodynamics and the Adler-Bardeen theorem for
the chiral anomaly in external electromagnetic fields. 
The chiral magnetic effect in hydrodynamics represents a macroscopic manifestation 
of a quantum phenomenon (chiral anomaly). Moreover, one can argue that
the current induced by the magnetic field
is  dissipation free and talk about a kind of "chiral superconductivity".
More precise description is a ballistic transport along magnetic field 
taking place in equilibrium and in absence of a driving force. 
The basic limitation is exact chiral limit while the temperature--excitingly enough-
does not seemingly matter.
 What is still lacking, is a detailed quantum microscopic
 picture for the ChME in hydrodynamics. Probably, the chiral currents
propagate through lower-dimensional defects, like vortices in superfluid. 
In case of superfluid, the prediction
  for the chiral magnetic effect remains unmodified
although the emerging dynamical picture differs
from the standard one. 
\end{abstract}

\section{Introduction}

In this chapter \footnote{The review is prepared for a volume of the Springer Lecture Notes in Physics  
"Strongly interacting matter in magnetic
fields" edited by D. Kharzeev, K. Landsteiner, A. Schmitt,
H.-U. Yee.}
we will consider chiral liquids, that is liquids whose
constituents are massless fermions. The motivation is an offspring from  the discovery of the 
strongly interacting quark-gluon plasma (for review see, e.g., \cite{shuryak})
which is, indeed,  build on (nearly) massless quarks.
The use of the (relativistic) hydrodynamic approximation is also suggested by the
observations  on the quark-gluon plasma. Moreover,  the state of the chiral liquid 
is assumed to be asymmetric with
respect to the left- and right-fermions. In other words, we concentrate on the
case of  a  non-vanishing chiral chemical potential $\mu_5$ \footnote{In the 
realistic QCD case 
the singlet axial current is anomalous and is not conserved.
Therefore, introduction of the chemical potential $\mu_5$ is rather a subtle issue.
In the bulk of the text we ignore this problem concentrating mostly
on academic issues. One could have in mind, for example,  that
the chemical potential $\mu_5\neq 0$ is associated in fact with the
axial current with isospin $\Delta I = 1$ which is conserved in the limit of vanishing
quark masses. Another possible line of reasoning is to invoke large-$N_c$ limit of 
Yang-Mills theories.
The contribution of the gluon anomaly is then suppressed by large $N_c$ 
and the chemical potential $\mu_5$ can be introduced consistently
for the singlet current as well.}.  
The motivation to introduce $\mu_5\neq 0$ is rather theoretical than experimental,
however, 
and is rooted in the 
sphaleron-based picture which predicts that, event-by-event, the plasma is produced as
chirally charged \cite{warringa}.

There are a few effects specific for the chiral liquids, the most famous one
being the chiral magnetic effect (for review and further references see
\cite{dunne}).
By the ChME one understands the phenomenon 
of induction of 
electromagnetic current  $\vec{j}_{el}$ by
applying an external magnetic field  $\vec{ B}$
to a chiral medium with a non-vanishing $\mu_5$ :
\begin{equation}\label{magnetic}
\vec{ j}_{el}~=~\frac{q^2\mu_5}{2\pi^2}\vec{ B}~,
\end{equation}
where $q$ is the electric charge of the constituents 
and $\mu_5, \mu_5=(\mu_R-\mu_L)/2$
is the chiral chemical potential.
Equation  (\ref{magnetic}) plays a central role in our discussion
and can be analyzed from various points of view. An exciting possibility
is that the chiral magnetic effect (\ref{magnetic}) has already been observed in
heavy-ion collisions, for a concise review and references see \cite{kharzeev5}.
We will concentrate, however, on the underlying theory rather than on 
its experimental verification.

Qualitatively, Eq (\ref{magnetic}) can be understood 
accounting only for the interaction of spin of quarks with an external magnetic
field, $H_s~\sim~q(\vec{\sigma} \cdot \vec{ B})$.
The overall coefficient in Eq. (\ref{magnetic}) 
 can  readily be found in case of free quarks \cite{vilenkin1,nielsen,dunne}.
Evaluation of the coefficient requires explicit counting of 
of zero modes of the Dirac equation for chiral fermions interacting with external magnetic
field \cite{nielsen}, see also Sect. 3.4.
The number of chiral zero
modes is controlled by
the famous chiral anomaly \cite{adler}.
Thus, Eq. (\ref{magnetic}) is a manifestation of the chiral anomaly,
as can actually be demonstrated in a number of ways,  discussed later.

One of central points is that the chiral magnetic effect can 
be derived not only in case of non-interacting chiral fermions but also in case
of strong  interactions between the constituents, provided
that the hydrodynamic approximation is granted. 
There is, though, a change in the Eq (\ref{magnetic}) which is of pure
kinematic nature. Namely, to describe liquid one introduces 
4-velocity of an element of the liquid, $u_{\mu}(x)$ which is a function of
the point $x$. The 4-velocity is normalized such that $-(u_0)^2+u_i^2=-1$,
and in the non-relativistic limit $u_0~\approx~1,~u_i~\approx~v_i$,
where $i=1,2,3$ and $\vec{v}$ is the 3-velocity entering  the
hydrodynamic equations in the non-relativistic limit. 
Eq. (\ref{magnetic}) is valid now 
only if the whole of the liquid is at rest.
If, on the other hand, $u_{\mu}$ is non-trivial Eq.
(\ref{magnetic}) is generalized to
\begin{equation}\label{magnetic1}
(j_{\mu})_{el}~=~{q^2\mu_5\over 2\pi^2}B_{\mu}~,
\end{equation} 
where $B_{\mu}~\equiv~1/2\epsilon_{\mu\nu\rho\sigma}
u_{\nu}(\partial_{\rho}A_{\sigma}-\partial_{\sigma}A_{\rho})$,
 $A_{\mu}$ is the gauge potential of the external electromagnetic field and the 
chemical potential $\mu_5$ can depend on the point $x$. In the rest frame of an element
of the liquid $u_{\mu}~\equiv~(1,0,0,0)$, and (\ref{magnetic1}) coincides with (\ref{magnetic}).

Non-renormalization theorems in field theory are quite exceptional and  attract 
a lot  of attention. Essentially, there are two best known  case studies  of non-renormalizability.
First, conserved charges are not renormalized, so that, for example, the absolute values 
of electric charges of electron and proton are the same. And the second example is 
the non-renormalizability of the chiral anomaly: 
\begin{equation}\label{anomaly}
\partial_{\mu}j^5_{\mu}~=~{\alpha_{el}\over 4\pi} \epsilon_{\alpha\beta\gamma\delta}
F^{\alpha\beta}F^{\gamma\delta}~~. 
\end{equation}
The absence of higher-order corrections to this anomaly is guaranteed by the
Adler-Bardeen theorem \cite{bardeen}.
 The proof of  non-renormalizability of the chiral magnetic effect
utilizes both symmetry considerations and the miracle of the perturbative cancellations,
revealed by the Adler-Bardeen theorem.

To put the consideration of the chiral magnetic effect into a field-theoretic framework
one considers hydrodynamics as a kind of effective field theory, see, e.g., \cite{son1}.
Viewed as an effective field theory, hydrodynamics reduces to expansion in a
number of
derivatives from the velocity $u_{\mu}$ and thermodynamic quantities. 
On the microscopic level, hydrodynamics corresponds to the  long-wave approximation, 
$$l/a~\gg~1~,$$ 
 where $l$ is of order of wave length 
of  hydrodynamic excitations and $a$ is of order distance between constituents.

The hydrodynamic equations
reflect symmetries of a dynamical problem considered since they 
are nothing else but the conservation laws.
In the absence of external fields
\begin{equation}\label{hydrodynamics}
\partial_{\mu}T^{\mu\nu}~=~0, ~~~\partial_{\mu}j^{\mu}_a~=~0~\end{equation} 
where $T_{\mu\nu}$ is the energy-momentum tensor and  $j^{\mu}_a$ 
are currents conserved in strong interactions, with the
index $a$ enumerating the currents.  
Consider liquid at rest and small fluctuations superimposed on it. Generically, fluctuations would
be  damped
 down on distances of order $a$ and do not propagate far away.  
 The exceptions   are fluctuations of conserved quantities which cannot disappear and
propagate far off. That is why the long-wave, or hydrodynamic approximation reduces to
the conservation laws (\ref{hydrodynamics}).

Explicit form of the hydrodynamic equations (\ref{hydrodynamics}) 
depends on how many terms in the gradient expansion are kept in  
$T^{\mu\nu}$ and $j^{\mu}_a$.
In general,
\begin{eqnarray}\label{tmunu}
T^{\mu\nu}=w u^{\mu} u^{\nu}+P g^{\mu\nu}+\tau^{\mu\nu}\\
j_a^{\mu}=n_a u^\mu+\nu_a^\mu,
\end{eqnarray}
where $w,P,n_a$ are the standard thermodynamical variables, namely, enthalpy, 
$w~\equiv~\epsilon+P$,  pressure and  densities of charges.
The quantities  $\tau^{\mu\nu}, \nu_a^\mu$ satisfy  conditions 
$u_\mu\tau^{\mu\nu}=u_\mu\nu_a^\mu=0$.
In the zeroth order in gradients $\tau^{\mu\nu}=0, \nu_a^\mu=0$.

The path from relativistic, or chiral hydrodynamics to the anomaly
(\ref{anomaly}) was found first in Ref. \cite{surowka}.
One introduces external electric and magnetic fields
so that the current conservation condition is changed into (\ref{anomaly}).
The bridge between a fundamental-theory equation (\ref{anomaly}) and
hydrodynamics is provided then by considering the entropy current $s_{\mu}$.
In presence of external fields the standard definition \cite{landauV} of the  current $s_{\mu}$
does not ensure the growth of the entropy any longer. To avoid the contradiction with
the second law of thermodynamics one includes terms proportional to external fields
both into the newly defined entropy current $s_{\mu}$ and currents $j^{\mu}_a$.
The constraints imposed by the second law of thermodynamics
involve the anomaly condition which is not renormalized by strong interactions
and turn to be strong enough to (almost uniquely) fix the currents  in
terms of the anomaly.   
We reproduce the basic point of this beautiful derivation in Sect.  2.1.

Most recently, it was observed \cite{general,jensen}
(see also \cite{cheianov}) 
that one can avoid considering the entropy current $s_{\mu}$.
Instead, one introduces not only external electromagnetic field but
 a static gravitational background as well.
Equating the hydrodynamic stress tensor and currents (\ref{tmunu})
to the corresponding structures evaluated at the equilibrium in the
gravitational background allows to fix the chiral current and stress tensor without
considering the entropy current.
This seems to be a very interesting extension of relativistic
hydrodynamics. From the perspectives of the present review, derivation \cite{general,jensen} 
reveals a novel feature of the chiral magnetic effect.
Namely, the corresponding electromagnetic current appears to be non-dissipative
since  it persists in the equilibrium. 
We will come back to discuss this point later.

All these derivations of the ChME in fact
uncover existence of some other effects as well. In particular, one predicts 
existence of the chiral vortical effect (ChVE),
namely, flow of axial current in the direction of local angular velocity
of the liquid, ${\vec{j}}^5\sim {\vec{\omega}}$.
In relativistic covariant notations:
\begin{eqnarray}\label{vortaic}
\delta j^5_\mu ~\approx~\frac{\mu^2}{2\pi^2}\omega_\mu \\ \nonumber
\omega_\mu = \frac{1}{2} \epsilon_{\mu\nu\alpha\beta} u^{\nu}\partial^\alpha u^\beta ,
\end{eqnarray}
where  $\omega_{\mu}$ is the vorticity of the liquid
and the chemical potential $\mu$ is considered to be small. The ChVE was derived first in a
holographic set up \cite{erdmenger} and is being actively discussed in the literature, along
with the chiral magnetic effect.  
Another example
is the axial-vector current $\vec{j}^5$ induced by a non-vanishing
 chemical potential $\mu_V$
\cite{son11,zhitnitsky}:
\begin{equation}\label{vector}
\vec{j}^5~=~\frac{q\mu_V}{2\pi^2}\vec{B}~~,
\end{equation}
which is a kind of parity-reflected companion of Eq. (\ref{magnetic}).

It is worth emphasizing that  all the chiral effects now considered 
were 
originally introduced quite long time ago basing on evaluation of loop graphs with
non-interacting fermions \cite{vilenkin,vilenkin1}. In particular, it was found 
in Ref. \cite{vilenkin} that rotating a system of non-interacting 
massless fermions results in a vortical current:
\begin{equation}\label{vilenkin}
\vec{j}^5~=~\Big({T^2\over 12}~+~{\mu^2\over 4\pi^2}\Big)\vec{\Omega}~,
\end{equation}
 where $\vec{\Omega}$ is the angular velocity of the rotation. In the term proportional 
to $\mu^2$ one readily recognizes Eq. (\ref{vortaic}) above 
\footnote{To compare (\ref{vilenkin}) 
and (\ref{vortaic}) one should keep in mind that in notations of Ref. \cite{vilenkin}
the current $\vec{j}$ in Eq (\ref{vilenkin}) is the current of right-handed fermions alone
and, thus, constitutes one half of the chiral current entering Eq. (\ref{vortaic}).
 }. It took many years, however, to prove that the result
 (\ref{vilenkin}) is essentially not modified by strong interactions.
As is mentioned above, the origin of the $\mu^2$ 
in Eq. (\ref{vilenkin}) term can be traced back
to the chiral anomaly  
and it is not renormalizable.

The status of radiative corrections to  the  $T^2$
term  in Eq. (\ref{vilenkin}) has been clarified only
very recently \cite{golkar,ren}. First, one relates
the chiral vortical effect  
to a static correlator of axial current and of momentum density.
As far as only fermionic part is kept in the operator of momentum  density,
all the higher-order contributions to the $T^2$ term cancel and result (\ref{vilenkin})
remains valid. The proof 
is based on analysis of Feynman graphs and echoes the
proof \cite{coleman} of
non-renormalizability  of the topological mass 
of a gauge field in 3d gauge theory.  There is, however, a  gluonic part
of the momentum density and it generates a calculable two-loop correction
to (\ref{vilenkin}).
  We reproduce the basic points of the proof  
following \cite{golkar} in Sect.  2.3.

Reference to anomalies of the fundamental theory
which arise due to weak coupling to external fields
(electromagnetic or gravitational) can be avoided by applying       
an effective field  theory.
 This effective field theory 
elevates chemical potentials to  interaction constants,
see  \cite{zahed,sadofyev,shevchenko} and references therein. 
The corresponding vertices can be obtained from the 
standard electromagnetic interaction by substitution
\begin{equation}\label{rule}
qA_{\mu}~\to~\mu u_{\mu}~~,
\end{equation}
where $\mu$ is the chemical potential associated with the conserved charge $q$.
The effective theory is anomalous and
 does reproduce through these anomalies the chiral magnetic effect
 and the $\mu^2$ term in the chiral vortical effect, see Eq. (\ref{vilenkin}).
\footnote{Note that in the underlying fundamental field theory
there are no anomalies associated with non-vanishing chemical potential $\mu$.
This observation is in no contradiction with the fact that such anomalies 
do arise in the language of
the effective theory.}. 
We will give further details in Sect. 2.4.

Non-renormalizability  commonly  implies topological nature
of the corresponding term. 
Moreover, if the currents are topological they are 
non-dissipative. The best known example of such a type is provided
by the integer quantum Hall effect (for the background and review see, e.g., \cite{hall}).
Consider a two-dimensional system with external electric field $\vec{E}=(E_1,0)$  applied.
Then there arises electric  current $j_i$ such that
$$j_i~=~\sigma_{ik}E_k~~,$$
where $\sigma_{ik}$ are coefficients.
 The integer quantum Hall effect is characterized by a non-diagonal $\sigma_{12}\neq 0$
 and $\sigma_{11}=0$:
\begin{equation}\label{hall}
\sigma_{12}~=~\nu{e^2\over h}~~, \end{equation}
where $\nu$ is integer. The work produced by the external electric field  is equal 
to the product $W~=~j_iE_i$. The Hall current is clearly not associated with any 
work done and this observation suggests strongly
that the Hall current is dissipation-less. 
For further examples of this type see, e.g., \cite{kimura}.

Since the chiral magnetic current (\ref{magnetic}) is not associated either
with any work done by an external field it seems natural to assume that the chiral
magnetic current is also dissipation-less.  This suggestion is made first in
Ref \cite{dissipation} basing on  somewhat different arguments, see  Sect. 3.1.

Now we come to a question, however, which has not been answered yet.
Namely,   topological, or dissipation-less currents
usually manifest existence of a macroscopic quantum state.
Well known examples are the superfluidity of weakly interacting Bose-liquid
or the same Hall current (\ref{hall}). In these two cases the nature of
the macroscopic quantum states
is well understood. Also, in case of non-interacting fermions
topological nature of the chiral magnetic effect has been demonstrated 
first long time ago \cite{nielsen}. Claiming the chiral magnetic current (\ref{magnetic})
to be topological in hydrodynamic approximation as well
 we imply  quantum nature of the corresponding ground state. 
 
  This problem of explicit constructing the quantum state can be addressed in 
some more detail within
   approach which starts with 
 a microscopical picture
and the central role  is played  then by low-dimensional defects.
A well known example of this kind is provided by
vortices in rotating superfluid.
In more abstract language,
 this approach
goes back also to papers   \cite{goldstonewilczec,nielsen,callanharvey}.
It was demonstrated that defects in field theory are   closely tied to the realization of anomaly.
In particular, it was shown in \cite{callanharvey} that anomaly in 2n+2 dimensional theory 
is connected with 2n dimensional index density and can be understood 
in terms of fermion zero modes on strings and domain walls.
In all the cases the chiral current is carried by fermionic zero modes
living on the defects.

 One can expect, therefore, that anomaly in effective, hydrodynamic 
  theory is realized in chiral superfluid system on vortex-like defects.  
The continuum-medium results (\ref{vortaic}), (\ref{magnetic})
can arise then upon averaging  over a large number of defects.
In case of the chiral magnetic effect   
such a mechanism was considered, in particular,  in Refs. \cite{zhitnitsky}, \cite{warringa}
 and
the final result (\ref{magnetic}) is reproduced on the microscopic level
as well. 
The vortices considered in \cite{zhitnitsky}, \cite{warringa} 
are simply the regions of space free of the medium substance.
In case of superfluidity the vortices are better understood
and the microscopical picture for the chiral effects can be clarified to some extent
\cite{kirilin}.
The outcome of the analysis in terms of defects, or vortices
is that the chiral magnetic effect does survive 
without any change.
As for the vortical chiral effect it is modified in the capillary picture
by a factor of two:
\begin{equation}
\Big(\delta J^5_\mu\Big)_{capillary} ~= ~2\frac{\mu^2}{2\pi^2}\omega_\mu 
\end{equation}
We will give details in Sect.  3.4.

Upon introducing the reader to the topics to be discussed in this review, 
we would like to emphasize that
there are many other  interesting results which could have been included into the review
but are actually not covered. The reason is mostly to avoid too much overlap with
other chapters  of this volume. A notable example of this kind
is the holographic approach to the ChME
which is reviewed, in particular, in Ref. \cite{landsteiner2}.
The same remark applies to the phenomenological manifestations of the gravitational anomaly. 
 Finally, there are very interesting applications
 of the technique used to condensed-matter systems.
 However, reviewing these applications goes beyond the scope of the present notes.
  
To summarize, we concentrate on two basic issues, non-renormalizability
and dissipation-free nature of the chiral magnetic and chiral vortical  effects.
In Sect. 2 we consider non-renormalization theorems  
 within various approaches outlined above
(thermodynamic, geometric, diagrammatic, effective field theories).
The derivations of  the theorems make it also clear that
the chiral effects considered are dissipation free.
In Sect. 3 we review further arguments in favor of dissipation-free
nature of the chiral effects. In this  section we also introduce
a microscopic picture in terms of defects  of  lower dimensions.
Sect. 4  is  conclusions.

\section{Non-renormalization Theorems}

\subsection{Non-renormalization theorems in thermodynamic approach}
In this subsection we reproduce the basic steps of the pioneering derivation 
\cite{surowka} of chiral effects, the chiral vortical effect first of all,
which utilizes only the chiral anomaly in 
external electromagnetic fields, thermodynamics and hydrodynamic
approximation. To simplify the algebra, 
we consider first a single conserved current, chiral at that.  Moreover, 
we consider the chiral symmetry not spontaneously broken (otherwise,
we should have modified hydrodynamics)

 In presence of external electromagnetic fields both the energy-momentum 
tensor and chiral current
are not conserved any longer.
 The current is not conserved because of the anomaly,
  while the
 energy is not conserved because external electric field executes work on the system.
 Thus, one starts with the equations:  
\begin{eqnarray}\label{nonconservation}
\partial_{\mu}j^{\mu}=CE^{\mu}B_{\mu},\\
\partial_{\mu}T^{\mu\nu}=F^{\nu\lambda}j_{\lambda}
\end{eqnarray}
where $C$ is the coefficient determined by the anomaly (e.g. for QED $C=\frac{1}{2\pi^2}$).
Turning to the thermodynamics, we have to introduce, following textbooks \cite{landauV},
an entropy current $s_{\mu}$ consistent with the second law of thermodynamics.
In the approximation of an ideal liquid  
the condition is $\partial_\mu s^\mu=0$ .

There are no general rules to construct $s_{\mu}$. 
As a first guess one can try $s_{\mu}=su_{\mu}$ where $s$ is the
entropy density. Moreover, put the gradient terms
$\nu^{\mu}, \tau^{\nu\mu}=0$ for simplicity.
 However, using the equations (\ref{nonconservation}) and $$dP=sdT+nd\mu$$
 where $\mu$ is chemical potential, one readily derives
\begin{eqnarray}
\partial_\mu (su^\mu)=-C\frac{\mu}{T}E\cdot B~~~~~~~~(\nu^{\mu},\tau^{\nu\mu}=0).
\end{eqnarray}
The right-hand side of this equation  does not have a definite sign 
and, therefore, one cannot accept $su^{\mu}$ as 
a definition of the entropy current in  presence of the anomaly.
Thus, we should continue with our guess-work to construct the entropy current.
Note that it is quite a common situation.
For example,  consider non-ideal liquid with non-zero $\nu^\mu$ (and $\tau^{\mu\nu}=0$). 
Then one has to modify \cite{landauV}  the entropy current
defining it as  $s^\mu=su^\mu-\frac{\mu}{T}\nu^\mu$ so that
 for the newly defined entropy current $\partial_\mu s^\mu=0$.

In presence of the chiral anomaly we can use the same idea and 
redefine the entropy current \cite{surowka} by  introducing terms proportional
to the magnetic field and vorticity. To simplify equations we will not
account for dissipative terms, viscosities and electrical conductance.
One can check that inclusion of these terms does not
change the result \cite{surowka}. Moreover,
in the next subsection we will see that there are general reasons for
the dissipative terms to be actually not relevant.  
Thus, expanding in the fields we look for solution for the matter current of the form:
\begin{eqnarray}
j_{\mu}~=~nu_{\mu}~+~\nu_{\mu}\\ \nonumber
\nu_{\mu}~=~\xi_{\omega}\omega_{\mu}+\xi_BB_{\mu}~~,
\end{eqnarray}
where $\omega_\mu=\frac{1}{2}\epsilon_{\mu\nu\alpha\beta}u^\nu\partial^\alpha u^\beta$ 
is the vorticity, 
$B_\mu=\frac{1}{2}\epsilon_{\mu\nu\alpha\beta}u^\nu F^{\alpha\beta}$ 
is the magnetic field in the rest frame of liquid element
(electric field $E_\mu=F_{\mu\nu}u^\nu$)  and $\xi_{\omega}, \xi_B$ 
are unknown functions of thermodynamic variables. 
For the entropy current, we assume:
\begin{eqnarray}\label{xiomega}
s_\mu=su_\mu-\frac{\mu}{T}\nu_\mu+D_{\omega}\omega_\mu+D_B B_\mu~,
\end{eqnarray}
where $D_{\omega}, D_B$ are further unknown functions.

Conservation of the entropy current now reads:
\begin{eqnarray}
\label{entropy0}
\partial_\mu \left(D_{\omega}\omega^\mu\right)+\partial_\mu \left(D_BB^\mu\right)-
\nu^\mu\left(\partial_\mu\frac{\mu}{T}-\frac{\mu}{T}\right)-C\frac{\mu}{T}E\cdot B=0.
\end{eqnarray}
For the ideal liquid ($\tau^{\mu\nu}=0$) the following identities hold:
\begin{eqnarray}
\partial_\mu\omega^\mu=-\frac{2}{w}\omega^\mu\left(\partial_\mu P-n E_\mu\right)\\
\partial_\mu B^\mu=-2\omega\cdot E+\frac{1}{w}\left(-B\cdot\partial P+n E\cdot B\right).
\end{eqnarray}
Moreover,   the coefficients in front of independent kinematical structures 
$\omega^\mu, B^\mu, E\cdot\omega, E\cdot B$ in relation 
(\ref{entropy0}) should vanish:
\begin{eqnarray}
\label{eqD1}
\partial_{\mu} D_{\omega}-2 \frac{\partial_{\mu}P}{\epsilon+p}D_{\omega}
-\xi_{\omega}\partial_{\mu}\frac{\mu}{T}=0\nonumber\\
\partial_{\mu} D_B- \frac{\partial_{\mu}P}{\epsilon+p}D_B-\xi_B\partial_{\mu}\frac{\mu}{T}=0\nonumber\\
\frac{2n D_{\omega}}{\epsilon+p}-2D_B+\frac{\xi_{\omega}}{T}=0\nonumber\\
\frac{n D_B}{\epsilon+p}+\frac{\xi_B}{T}-C\frac{\mu}{T}=0.
\end{eqnarray}
To proceed further one has to choose a basis 
of thermodynamic variables and it is convenient to take this basis as 
$\left(P,\tilde{\mu}=\frac{\mu}{T}\right)$.
 The thermodynamic derivatives in the basis look as
$\left(\frac{\partial T}{\partial P}\right)_{\tilde{\mu}}=\frac{T}{w}, 
\left(\frac{\partial T}{\partial\tilde\mu}\right)_{\tilde{P}}=
-\frac{nT^2}{w}$ . Since  the  thermodynamic 
gradients $\partial_\mu P, \partial_\mu \tilde{\mu}$
are independent the first two equations in (\ref{eqD1})
imply four conditions: 
\begin{eqnarray}
\label{eqD2}
-\xi_{\omega}+\frac{\partial D_{\omega}}{\partial\tilde\mu}=0, ~~~
-\xi_B+\frac{\partial D_B}{\partial\tilde\mu}=0\nonumber\\
\frac{\partial D_{\omega}}{\partial P}-\frac{2}{w}D_{\omega}=0,~~~~
\frac{\partial D_B}{\partial P}-\frac{1}{w}D_B=0,
\end{eqnarray}
and the  general solution for $D_{\omega},D_B$ looks as 
$$D_{\omega}=T^2d_{\omega}(\tilde\mu), ~~~D_B=Td_B(\tilde\mu)~~,$$ 
with functions $d_{\omega},d_B$
being so far arbitrary. Then the last two equations in (\ref{eqD1}) reduce 
to simple differential equations which can be readily solved.
As a result, the functions $d_{\omega}(\bar{\mu}),d_B(\bar{\mu})$ get
fixed, up to the integration constants \cite{neiman,sadofyev}
which are the values of the functions
at $\bar{\mu}=0$.  
 
For the  chiral kinetic coefficients we finally obtain:
\begin{equation}
\label{coeff1}
\xi_{\omega}~=~C\mu^2(1-\frac{2}{3}\frac{\mu\cdot n}{\epsilon+p})~+
~C_{\omega}T^2(1-\frac{\mu\cdot n}{\epsilon+p})
\end{equation}
and 
\begin{equation}\label{coeff11}
\xi_B~=~C(\mu-\frac{1}{2}\frac{\mu^2 n}{\epsilon+p}),
\end{equation}  
where the constant $C$ determines the anomaly and is fixed while
the constant $C_{\omega}$ remains undetermined. 
We will come back to evaluate $C_{\omega}$
in Sect. 2.3  following the paper \cite{golkar}. Note that  we omitted a
similar 
constant of integration from (\ref{coeff11}).  The reason is that
such a constant  can appear in fact
 only due to parity-violating interactions \cite{neiman} and, having in mind eventual
applications to parity-conserving theories, we suppressed it right away. 

The non-vanishing 
$\xi_{\omega}, \xi_B$ exhibit what we call chiral vortical and chiral magnetic effects, respectively.
At temperature $T=0$ the functions $\xi_{\omega}(\mu),\xi_B(\mu)$
are fixed in terms of the coefficient $C$ which can be read off from the chiral anomaly
and is not renormalized by strong interactions. This is the content of the
non-renormalization theorem of the chiral effects in hydrodynamic approximation.
It is worth emphasizing that if parity
is conserved the chiral magnetic and  vortical effects are manifested in fact in different currents,
axial and vector, respectively. On the other hand, according to Eq (\ref{xiomega}), 
they appear in one and the same current.
The reason is that through postulating conservation of
a single chiral current we actually admitted for a parity violating "strong interaction".
The case of a few currents, which allows for parity-conserving strong interactions
was considered, in particular in Refs \cite{neiman,sadofyev}.
The outcome of the calculation is essentially the same: the chiral effects are fixed in all
the currents, up to  constants of integration.

\subsection{Non-renormalization theorems in geometric approach}

Let us recall the reader one of simplest derivations of the chiral magnetic 
effect \cite{nielsen,warringa}. The anomaly,
\begin{equation}
\partial^{\mu}j_{\mu}^5~=~{q^2\over 2\pi^2}\vec{ E}\cdot\vec{B}
\end{equation}
can be rewritten as an equation for the production rate of chiral 
particles. Denoting the total chirality as $N_5$, where $N_5\equiv N_R-N_L$
and $N_R (N_L)$ is the number of right- (left-) handed particles
we have:
\begin{equation}\label{tobereplaced}
{dN_5\over dt d^3x}~=~{q^2\over 2\pi^2}\vec{ E}\cdot\vec{B}~~.
\end{equation}
Production of particles requires for energy to be deposit into the system.
The source of this energy is the work done by the
external electric field. Therefore: 
\begin{equation}\label{linear}
\int d^3x \vec{j}_{el}\cdot \vec{E}~=~\mu_5{dN_5\over dt}~=
~{q^2\mu_5\over 2\pi^2}\int d^3x {\bf B}\cdot \vec{ E}
\end{equation}
where $\vec{ j}_{el}$ is the electric current and $\mu_5$ is the energy 
needed to produce a particle.
Tending $\vec{ E}\to 0$ we learn from Eq. (\ref{linear})  
that there survives a non-vanishing
current in this limit: $\vec {j}_{el}~=~(q^2\mu_5/2\pi^2)\vec{ B},
$ and we come back to Eq. (\ref{magnetic})
As is mentioned above, the current is non-dissipative  
since  magnetic field does not produce any work.
One can also say that  the current $\vec{j}_{el}$  exists in the equilibrium.

To summarize, one can calculate the non-dissipative current associated with the magnetic field
by introducing electric field, taking the system out of the equilibrium in this way
and then tending the electric field
back to zero.
A similar technique is commonly applied  to study spontaneous symmetry breaking.

Recently it has been realized \cite{general,jensen} that the procedure
can be generalized in a rather unexpected way. Namely, one introduces
not only external electromagnetic field but static gravitational field as well
and studies equilibrium  in this background. All the terms in chiral currents and 
energy-momentum tensor are fixed in the equilibrium and are non-dissipative.
Eventually one  can can go  back to the flat space.

The basic object in the approach \cite{general,jensen} is the generating functional $W$
as a function of external electromagnetic and gravitational fields, or sources,
$$W~=~\int d^dxL(sources (x))~~,$$
where $W=\ln Z$ and $Z$ is the partition function.
Differentiating $W$ with respect to the sources one evaluates in the standard way
energy-momentum tensor and currents, as well as and their correlators
in the equilibrium. In particular,
\begin{equation} \label{differentiate}
<T_{\mu\nu}>~=~{2\over \sqrt{-g}}{\delta~W\over \delta g_{\mu\nu}}~,~~
<j_{\mu}>~=~{1\over \sqrt{-g}}
{\delta~W\over \delta A_{\mu}}.
\end{equation}
In the spirit of the hydrodynamic approximation, one expands $W$ in
the number of derivatives, both from
the sources and thermodynamic variables and reiterates the procedure
in each order in the expansion.

The medium is characterized by the time-like vector $u^{\mu}$ which in the zeroth order
in the number of derivatives can be chosen as $u^{\mu}~\sim~(1,0,0,0)$. 
Apart from $u^{\mu}$ the generating functional can depend on observables
that are local in space but non-local in Euclidean time.
In the zeroth order in derivatives, 
 the invariant length $L$ of the time circle is one  of such
 observables. Also, there is  Polyakov loops $P_A$ of any
U(1) gauge fields. Therefore, temperature $T$ and chemical potential $\mu$ are
defined geometrically as
\begin{equation}\label{invariant}
T~=~{1/ L}~,~~\mu~=~\ln P_A/L~.
\end{equation} 
These are simplest examples of diffeomorphic and gauge invariant scalars.

We pause here to emphasize that the outcome of the calculation are static correlators
which are the same in the Euclidean and Minkowskian versions
of the theory. Therefore the relations obtained are in fact thermodynamic in nature.
Static correlators are to be distinguished from correlators
which determine, through the Kubo formula, such transport coefficients as viscosity.
In the latter case one considers correlator of certain components of  the stress tensor 
at momentum transfer $\vec{ q}~\equiv~0$ and frequency
$\omega~\to~0$. On the other hand, correlators 
considered here correspond to $\omega~\equiv~0$, $\vec{q}~\to~0$.
This, subtle at first sight, difference is crucial for continuation
to the Euclidean space.
It is straightforward
to realize that the chiral magnetic effect for a time-independent magnetic field is
indeed determined by a static correlator of components of electromagnetic current,
see, e.g., 
\cite{kharzeev7,landsteiner1}. To demonstrate this, it is convenient to
begin with the standard Kubo relation for electric
conductance:  
\begin{equation}
\sigma_E~=~\lim_{\omega\to 0}{{i\over \omega}<j_i,j_i>|_{\vec{q}=0}}
\end{equation}
where $\sigma_E$ determines the electric current in terms of a time-independent electric field,
$\vec{ j}_{el}~=~\sigma_E\vec{E}$, and $<(j_{el})_i,(j_{el})_i>$
is the retarded correlator of the components of the electromagnetic current 
(with no summation over the index $i$).
 Since both electric field $\vec{ E}$ and magnetic field $\vec{B}$ are related 
 to the same vector-potential $\vec{A}$
($\vec{ E}=-i\omega \vec{A}$, $\vec{B}=i{\bf q}\times \vec{A}$)
one concludes:
\begin{equation}\label{sigmab}
\sigma_B~=~\lim_{q_n\to 0}\sum_{ij}{\epsilon_{ijn}{i\over 2q_n}
<({ j}_{el})_i,({ j}_{el})_j>|_{\omega=0}}~,
\end{equation}  
where $\sigma_B$ is defined as $\vec{j}_{el}~=~\sigma_B\vec{B}$.
Thus, it is indeed a static correlator which we need to evaluate the ChME.

Probably, the best known example of the use of static correlators is the
generation of photon screening mass through the Higgs mechanism.
Namely, the
correlator of components of electromagnetic current in superconducting case
looks as:
$$\lim_{\vec{q}\to 0}{\int d^3x~\exp(i\vec{q\cdot r})~<j_i(\vec{x}), j_k({0})>}
~\sim~\big(\delta_{ik}-\vec{q}_i\vec{q}_k/\vec{q}^2\big)~,$$ 
where presence of a pole signals superconductivity while
 the local term proportional to $\delta_{ik}$ signifies
a non-vanishing photon mass.
A similar role of a signature of superfluidity is played 
by  a pole in the static correlator
of the components of momentum density:
\begin{equation}\label{3d}
\lim_{\vec{q}\to 0}{\int d^3x~\exp(i\vec{q\cdot r})~<T_{0i}(\vec{x}), T_{0k}({0})>}
~\sim~\vec{q}_i\vec{q}_k/\vec{q}^2~~.
\end{equation}
Note, however, absence of the local term proportional to $\delta_{ik}$.
This result is readily understood if we start from considering non-trivial 
gravitational background.
The local terms are associated then with  covariant derivatives, say,
$D_iv_k~=~\partial_iv_k+\Gamma_{ik}^lv_l$, where $v_i$ is a vector
and $\Gamma_{ik}^l$ are the Christoffel symbols. The $\Gamma_{ik}^l$ symbols contain 
only derivatives from the components of the metric tensor $g_{\mu\nu}$
and we immediately conclude
that there could be no $\delta_{ik}$ local term in the correlator of $T_{0i}$ components.
Thus, we see that introducing first gravitational background does allow 
to fix the subtraction term in a static correlator in the flat space.

Currents which we are considering now are somewhat similar to the
standard superfluid current \cite{landau9}.
But there are important differences as well. In particular, the value of the superfluid current is not
 fixed in the equilibrium while the current associated with 
 the ChME has a unique value. 
Now, the statement is \cite{general,jensen} that all the non-dissipative 
pieces in $<j_{\mu}>,~<T_{\mu\nu}>$  can be 
conveniently determined by embedding the system into  static 
electromagnetic plus  gravitational
background. 
 We will outline briefly the proof following \cite{general}. 

Static gravitational background in all the generality
can be parameterized as follows:  
\begin{equation}\label{metric}
ds^2~=~-e^{2\sigma(x)}\big(dt+a_i(x)dx^i\big)^2+g_{ij}(x)dx^idx^j~~,
\end{equation}
where $x_i$ are spatial coordinates ($i=1,2,3$ for definiteness),  $\partial_t$ 
is the Killing vector on this manifold, gravitational potentials
 $\sigma,a_i,g_{ij}$ are smooth functions of
the coordinates $x_i$. One assumes also presence of a static U(1) gauge field ${\it A}$,
\begin{equation}
{\it A}~=~{\it A}_0dx^0+{\it A}_idx^i
\end{equation} 
The ${\it A}_0$ component is related to the chemical potential, see Eq (\ref{invariant}).

Consider first zeroth order in gradients. Then it is quite obvious that in the
equilibrium
\begin{equation}
u^{\mu}_{(0)}~=~e^{-\sigma(x)}(1,0,0,0),~~~T_{(0)}(x)~=~e^{-\sigma(x)}T_0,
~~~\mu_{(0)}(x)~=~e^{-\sigma(x)}{\it A}_0~,
\end{equation}
 where $\sigma(x)$ enters the metric (\ref{metric}) and subscript (0) refers to the zeroth
order in expansion in derivatives. Indeed, expressions for $T_{(0)}(x), \mu_{(0)}(x)$ 
can be obtained
directly from their invariant definition (\ref{invariant}) while $u^{\mu}_{(0)}(x)$
is fixed by the normalization condition.
Thus, the function $W$ to this lowest order 
reduces to
\begin{equation}
W_{(0)}~=~\int \sqrt{g_3}{e^{\sigma}\over T_0}P(T_0e^{-\sigma},{\it A}_0e^{-\sigma})~,
\end{equation}
where $P(T,\mu)$ is pressure as function of temperature and chemical potential in flat space.

In higher orders in derivatives one expands $u^{\mu},T,\mu$ further:
\begin{eqnarray}
u^{\mu}~=~u^{\mu}_{(0)}~+~u^{\mu}_{(1)}~+~u^{\mu}_{(2)}~+~...~~,\\
T~=~T_{(0)}~+~T_{(1)}~+~T_{(2)}~+~...~~,\\
\mu~=~\mu_{(0)}~+~\mu_{(1)}~+~\mu_{(2)}~+~...~~,
\end{eqnarray}
where $u^{\mu}_{(n)},T_{(n)}, \mu_{(n)}$ are expressions of n-th order in derivatives acting on 
the background fields
$\sigma, {\it A_0},{\it A}_i,g_{ij}$. It is important that both $T_{(n)}$ and $\mu_{(n)}$
are constructed on the same set of gauge- and diffeomorphic-invariant scalars. 
Let us denote the
number of such scalars  as $s_{(n)}$. As for the four velocity $u^{\mu}$
normalized to unit,
 $\delta u^{0}$ can be expressed
in terms of $\delta u^i$ and is not independent. Variations of the vector $u^i_{(n)}$
are expanded in the set of independent invariant vector combinations. 
The total number of such combinations is denoted as $v_{(n)}$.

Consider now a
general decomposition of the energy-momentum tensor and of the current $j_{\mu}$:
\begin{eqnarray}
T^{\mu\nu}~=~{\it E}u^{\mu}u^{\nu}+{\it P}\Delta^{\mu\nu}+q^{\mu}u^{\nu}+q^{\nu}u^{\mu}+
\tilde{\tau}^{\mu\nu},\\
j^{\mu}~=~{\it N}u^{\mu}+\nu^{\mu}
 \end{eqnarray}
where $q^{\mu}u_{\mu}=\tilde{\tau}^{\mu\nu}u_{\mu}=\Delta^{\mu\nu}u_{\mu}=0$,
$g_{\mu\nu}\tilde{\tau}^{\mu\nu}=0$. Similar to the observations above, the scalars ${\it E,P,N}$
can be written as expansions in independent gauge- and diffeomorphic-invariant scalars,
vector $q^{\mu}$ is expanded in independent  SO(3) vector structures while expansion
of $\tau^{\mu\nu}$ would require introduction  of a set of independent SO(3) tensor
structures, with trace zero. The total number of the tensor structures is denoted by $t_{(n)}$.
Note that within the standard relativistic hydrodynamics the energy-momentum tensor is defined
somewhat different, in the so called Landau gauge, see Eq (\ref{tmunu}). 
Namely, there is no vector $q^{\mu}$
and $\epsilon,P, n$ are those functions of temperature $T$ and of chemical potential
as determined by flat-space equilibrium thermodynamics. As a result, it is only $\tau_{\mu\nu}$
and $\nu^{\mu}$ which are to be expanded  in gauge and diffeomorphic invariant structures
introduced above. Therefore, the hydrodynamic tensors (\ref{tmunu}) are expanded in
$s_{(n)}+v_{(n)}+t_{(n)}$ invariant structures.

The central point of the procedure invented in \cite{general,jensen} is to equate 
the standard hydrodynamic tensors (\ref{tmunu}) to the expressions
obtained by differentiating the functional $W$, see Eq (\ref{differentiate}).
This can be done only at the equilibrium point. In the equilibrium, not all the
gauge- and diffeomorphic-invariant structures introduced above survive. The number
of non-vanishing structures is called $s^e_{(n},v^e_{(n)},t^e_{(n)}$. Moreover,
these, surviving terms are non-dissipative since they exist in the equilibrium.
As for the dissipative terms, which correspond to the structures vanishing
in the equilibrium one gets no predictions or constraints concerning them.

By simple counting, the total number of the constraints is $3s^e_{(n)}+2v^e_{(n)}+t^e_{(n)}$
and this number is exactly the same 
as needed to both find expansion corrections of the n-th order to $T,\mu,u^i$  
and determine the equilibrium stress tensor and current in the Landau gauge
(\ref{tmunu}) \footnote{Alternatively, this counting can be 
considered as a proof of the possibility
to introduce the Landau gauge (\ref{tmunu}).}. 
Indeed, the expansion of $T,\mu,u^i$ at the equilibrium point  depends on 
$2s^e_{(n)}+v^e_{(n)}$
parameters while expansion of the energy-momentum tensor and of the current in
the Landau gauge brings in $s^e_{(n)}+v^e_{(n)}+t^e_{(n)}$ terms.
This completes the general proof that all the non-dissipative terms,
like the chiral magnetic effect, are fixed uniquely. Actual details and explicit examples
can be found in Refs \cite{general,jensen}.   In particular, the results of Ref \cite{surowka}
have been reproduced. 

The method of Refs. \cite{general,jensen} outlined above allows to derive systematically 
chiral effects, like the ChME, for any number of conserved and anomalous currents
and to any order in the derivative expansion. Remarkably, one avoids   
considering the entropy current $s_{\mu}$ altogether. The derivation makes it clear that
one can fix only currents existing in the equilibrium. In  other words, the currents 
are non-dissipative
\cite{landsteiner1}.
The issue will become central for us in Section  3.

\subsection{Non-renormalization theorems in diagrammatic approach}

Very recently, there was a remarkable development \cite{golkar,yarom} in understanding the
temperature-dependent chiral vortical effect, see the $T^2$ term in Eqs. (\ref{vilenkin}),
(\ref{coeff1}).
Namely, it was demonstrated that the bare-loop result (\ref{vilenkin})  
is modified in two-loop order in a well defined way.  

As a preliminary remark, let us notice that the chiral vortical effect is determined
\cite{landsteiner1} in terms of a
static correlator, similar to the case of the chiral magnetic effect, see Eq. (\ref{sigmab}).
We recall the reader that the chiral vortical effect is defined in terms of the coefficient
$\xi_{\omega}$, see Eq. (\ref{xiomega}). In the non-relativistic limit we have 
the following piece in the axial-vector  $j^5_i$:
\begin{equation}\label{definition}
\delta ( j^5_i)~=~\xi_{\omega}\epsilon_{ijk}\partial_jv_k~~,
\end{equation}
where $v_k$ are components of the 3-velocity of an element of the liquid.
Then for the coefficient $\xi_{\omega}$ one gets \cite{landsteiner1}:
\begin{equation}\label{xiomega}
\xi_{\omega}~=~\lim_{q_n\to 0}\sum_{ij}{\epsilon_{ijn}{i\over 2q_n}<j^5_i,T_{0j}>|_{\omega=0}}~~,
\end{equation}    
(no summation over index $n$).
The argumentation is based on the well known analogy between the vector 
potential of a gauge field $\vec{A}$ and the metric-related vector $\vec{g}$,
where $g_i\equiv~g_{0i}$, or $ds^2=dt^2+2g_idtdx^i+dx_i^2$.
In the rest-frame of the fluid but in the background of the gravitational
potential $\vec{g}$ we have for the 4-velocity of the liquid $u_{\mu}= (-1, \vec{v}) =
(-1, \vec{g})$. Therefore the "gravi-magnetic field"  $\vec{B}_g~\equiv~curl \vec{g}$
and we find, indeed, a complete analogy between the coefficient $\sigma_B$, 
see Eq. (\ref{sigmab}), 
describing the chiral magnetic effect  and the coefficient  $\xi_{\omega}$,
see Eqs. (\ref{definition}) and (\ref{xiomega}), 
describing the "chiral gravi-magnetic effect", or the chiral
vortical effect as we call it.   

Let us recall the reader that the non-renormalization theorem of Sect. 3.1 fixes
the chiral vortical effect up to a temperature -dependent term, proportional to $T^2$
which can be evaluated at the vanishing chemical potential $\mu=0$, see Eq.(\ref{coeff1}).
We turn  now to the problem of evaluating this missing term.
In view of Eq. (\ref{xiomega}) we are interested then in the following term in
the effective action:
\begin{equation}\label{seffective}
S_{eff}~=~i\xi_{\omega}\int d^3x\epsilon_{ijk}A^5_i\partial_j g_k~\equiv~iT^2C_{\omega}
\int d^4xa_i\partial_jb_k~,
\end{equation}
where  $a_i\equiv A^5_i$ is the gauge field coupled to $j^5_i$, $Tb_i~\equiv ~ g_{0i}$ is a component of
the metric. We consider linearized gravity and the action is to be invariant under
gauge and diffeomorphic transformations
\begin{equation}\label{shifts}
a_i~\to~a_i~+~\partial_i\alpha,~~~b_i~\to~T(\nabla_i{\epsilon_0}+\nabla_0\epsilon_i)
\end{equation}
where $\alpha$ is an abitrary function, $\epsilon_{\mu}$ is the diffeomorphism parameter,
$x_{\mu}\to x_{\mu}+\epsilon_{\mu}$. 

The Lagrangian density corresponding to the action ({\ref{seffective}) describes a 3d
non-diagonal mass term mixing wave functions of the vectors $a_i$ and $b_i$.
The action (\ref{seffective}) is gauge and diffeomorphic invariant, as it should be.
However the mass term itself is not. This  observation allows eventually
to prove cancellation of a large class of radiative corrections \cite{golkar}. 
The 3d nature of the action (\ref{seffective}) is crucial  for the proof.
 Note that although we started with a 4d gauge theory,  
reduction to a sum over a sequence of 3d theories
is inherent to the problem since at  finite temperature any
   4d field theory reduces to 
 a sum over Matsubara frequencies, with
each frequency corresponding to a 3d theory.

The analysis of the radiative corrections to  the 3d topological term (\ref{seffective})
echoes the proof of non-renormalizability of a pure gauge-boson topological mass
given about 30 years ago \cite{coleman}. In that case the topological 
mass looks as \cite{jackiw}:
\begin{equation}\label{topmass}
S^{gauge}_{eff}~=~im_g\int d^3x \epsilon_{ijk}a_i\partial_ja_k~~.
\end{equation}
 This topological mass arises on one-loop level in 3d gauge theories. 
The simplest Lagrangian of the
matter field looks as
\begin{equation}\label{mlagrangian}
L_m~=~\bar{\psi}(D_{\mu}\gamma_{\mu}-m_0)\psi~,
\end{equation}
where $D_{\mu} (\mu=0,1,2)$ are covariant derivatives. The one-loop contribution
is given by
\begin{equation}\label{oneloop}
(m_g)_{one-loop}~=~{q^2\over 4\pi}{m_0\over |m_0|}~~,
\end{equation}
where $q$ is the charge of the fermion.
In higher orders of perturbation theory one-loop fermion graph with a few
photon exchanges arise. One can integrate first over the (massive) fermion
and reduce the graph to a n-photon effective vertex. In the momentum space, 
it is denoted as
$\Gamma^{(n)}_{\mu_1...\mu_n}(q_1,...,q_n)$ where $q_i$ are photon 
momenta and the overall
$\delta$-function is factored out. The vertex is a function of $(n-1)$ independent momenta,
$q_1,...,q_{n-1}$. Take now the limit of all the momenta $(q_1,...,q_{(n-1)})$ small.
Then it is straightforward to prove that the effective vertex is to vanish in the limit of any
independent momentum zero. In other words,
\begin{equation}\label{vanishing}
\Gamma^{(n)}(q_1,...,q_n)~=~ O(q_1\cdot q_2\cdot...\cdot q_{n-1})~.
\end{equation}
This relation is sufficient to prove that all the contributions to topological mass, 
beginning with two loops, vanish.
Indeed, one can always choose the momenta corresponding to the external photon legs
to be included into momenta $(q_1,...,q_{(n-1)})$
which are small. The one-loop graph is exceptional in this sense since there is
only a single independent photon momentum and
$\Gamma^{(1)}(q_1)~=~O(q_1)$. 

Let us come back to discussion of the topological mass (\ref{seffective}) relevant to 
the vortical effect. It is determined by the correlator (\ref{xiomega}).
Let us split the momentum-density operator into fermionic and gluonic parts:
\begin{equation}\label{split}
T_{0j}~=~(T_{0j})_{fermionic}+(T_{0j})_{gluonic}~.
\end{equation}
As far as we keep  only the fermionic part
 the main idea, that only 3d one-loop graphs
can contribute to (\ref{seffective}), remains the same. There is, however, an important change
concerning infrared behavior of higher-loop graphs. 
Amusingly, the masslessness of the fermion in the original 4d field theory
does not matter
since in the 3d projection the fermions do have non-vanishing masses,
$m_f^2~=~4\pi^2(n+1/2)^2$ where $n, ~(n=0,1,2...)$ enumerates Matsubara
frequencies.
The actual subtle point is that
in non-Abelian 3d theory higher loops generically diverge badly 
in the infrared.  
According to Ref. \cite{golkar} the infrared cut off 
is still provided by the non-perturbative gluon mass emerging \cite{gross} at finite
temperatures. The effective gluon mass is of order $(m^2_g)_{eff}\sim g^4_{Y-M}(T)T^2$
and at external momenta much smaller than this effective mass one can expect
that higher-loop contribution to the topological mass still vanishes.
Finally, the evaluation of one-loop fermionic 
contributions to the topological mass (\ref{seffective})
is more involved technically
than in case of (\ref{topmass}). Each 3d theory
corresponding to the n-th Matsubara frequency does contribute to $(\ref{seffective})$.
Indeed, according to (\ref{oneloop}) the one-loop contribution does not 
disappear with the growing fermion mass.
As a result, one comes \cite{golkar} to a divergent sum.
The standard $\zeta$-function regularization
provides the final answer for the temperature-dependent vortical effect:
\begin{equation}\label{finall}
C^{ferm}_{\omega}T^2~=~-T^2\sum\limits_{m=1}^{\infty}m~\to~{1\over 12}T^2~~,
\end{equation} 
where $C^{ferm}_{\omega}$ is the contribution of
the fermionic loop  to $C_{\omega}$.

So far we discussed fermionic part of $T_{0j}$, see 
(\ref{split}) and argued that its contribution is exhausted by one-loop 
graphs. The argument does not apply, however, to the gauge-field part, $(T_{0j})_{gluonic}$.
It does generate a calculable two-loop radiative correction to the $T^2$ 
term in Eq. (\ref{vilenkin}),
see \cite{ren,golkar}.
 We should not feel disappointed about this lack of complete
cancellation of higher loops. Indeed, although the Adler-Bardeen theorem is
commonly referred to as a proof of non-renormalization of the anomaly,
it is known since long (see, e.g., \cite{anselm}) that two-loop graphs corresponding
to rescattering of gauge field do not vanish in fact. The proof of non-renormalizability
of the ChME in Sect. 3.1 avoided this problem only because we treated
the electromagnetic field as external (not dynamical).
Note also that the $T^2$ term in the chiral vortical effect
 is present only in case of singlet currents
which are anomalous anyhow and the status of the ChME is not so clear. 
Its evaluation, however, is an amusing demonstration that dissipation-free
processes generically are  not suppressed at all   
at high temperature.

\subsection{Non-renormalization theorems in effective field theories}
 
Here we develop another approach to the chiral effects based on an effective theory
following  mostly Ref. \cite{shevchenko}.
The basic idea is to treat chemical potentials as  effective couplings.
The basic rule can be memorized as an analogy between interaction
with charge $q$   and with chemical potential $\mu$:
$
qA^{\mu}~\to~\mu u^{\mu}$,
where $A_{\mu}$ is the vector-potential of an external field and $u^{\mu}$ is, as usual, 
the 4-velocity of an element of  liquid, see Eq. (\ref{rule}).

Let us first substantiate the rule (\ref{rule}).
 Chemical potentials are   introduced  through the effective Hamiltonian:
\begin{equation}
\label{ui}
\delta H = \mu Q + \mu_5 Q_5,
\end{equation}
where $Q=\int d^3 x \psi^\dagger\psi$ and $Q_5=\int d^3 x \psi^\dagger \gamma_5\psi$.
Note that this is indeed an effective, not fundamental interaction since any chemical 
potential is introduced thermodynamically, that is for  a large number of particles. 
However, at least formally the Hamiltonian (\ref{ui}) looks exactly the same as the Hamiltonian for 
a fundamental  (i.e., not effective) interaction of charges with the $A_0$ gauge field.
Thus, we come to the analogy (\ref{rule}) in the particular case $u^0=1, u^i=0$.
This case was considered in many papers, see in particular \cite{son1}, and 
in the equilibrium the analogy between
$A_0$ and $\mu$ is commonly used nowadays. We have already exploited this connection,
see Eq. (\ref{invariant}). In this sense the effective theory considered
here can be viewed as a simplified version of a much more elaborated
scheme outlined in Sect.  2.2.
 
 Treating (\ref{ui}) as a small perturbation we can also fix the corresponding  Lagrangian density: 
\begin{eqnarray}\label{effective}
S_{eff}=\int dx\left(i\bar{\psi}\gamma^{\rho}D_{\rho}\psi+\mu\bar{\psi}\gamma^{0}\psi
+\mu_5\bar{\psi}\gamma^{0}\gamma_5 \psi\right)+S_{int},
\end{eqnarray} 
where $D_{\mu}$ is the covariant derivative in external electromagnetic field and $S_{int}$ is a
fundamental interaction responsible for the formation of the liquid. For our discussion it is
 crucial that  $S_{int}$ does not induce any anomaly. We can  drop then this  interaction
for our purposes.

 So far we considered the whole of the liquid as being at rest
 and the chemical potential being constant through the whole of the volume. 
 To study hydrodynamics one can use the standard trick of  boosting the action
 into a local rest frame by utilizing the 4-vector $u^{\mu}$:
\begin{eqnarray}\label{relativistic}
S_{eff}=\int dx\left(i\bar{\psi}\gamma^{\rho}D_{\rho}\psi+\mu u_\mu\bar{\psi}\gamma^{\mu}\psi
+\mu_5u_\mu\bar{\psi}\gamma^{\mu}\gamma_5 \psi\right)~.
\end{eqnarray}
The Lagrangian density (\ref{relativistic}) 
coincides in the rest frame with (\ref{effective})
but looks perfectly Lorentz invariant and can be used in  covariant perturbative calculations.
The boost velocity $u^{\mu}$ is treated then as 
a slowly varying external field, similar to $A^{\mu}$.

  The only difference  between the effective action (\ref{relativistic}) and  fundamental one 
  is the presence of the terms proportional to the chemical potentials
  $\mu_{5} , \mu$.  The fundamental theory is free 
  from anomalies in the case considered now, $(E\cdot  B)=0$.
  Therefore the only possible source of anomalies in the  effective theory
  are the triangle graphs with $\mu,\mu_5$ entering the vertices.
  
  The presence of an anomaly in the effective theory can indeed be readily verified. 
  For this purpose 
one can calculate triangle graphs or use Fujikawa-Vergeles path-integral considerations. 
According to the latter technique, anomalies emerge due to non-invariance of the path-integral 
measure under field transformations. Consider the following transformation
\begin{equation}
\psi\rightarrow e^{i\alpha\gamma_5+i\beta}\psi~.
\end{equation}
Then, by the standard technique
one readily  finds \footnote{We could have defined anomaly in such a way that it does not contribute
to $\partial_{\mu}j^{\mu}$.  However, in the presence of both chemical potentials  $\mu$ and $\mu_5$ there is no
physical motivation for such a regularization.}:
\begin{eqnarray}
\label{anomaly1}
\partial_\mu j_5^{\mu}=-\frac{1}{4\pi^2}\epsilon_{\mu\nu\alpha\beta}
\left(\partial^\mu(A^\nu+\mu u^\nu)\partial^\alpha (A^\beta+ \mu u^\beta)+
\partial^\mu\mu_5 u^\nu\partial^\alpha \mu_5 u^\beta\right)\\
\label{anomaly2}
\partial_\mu j^{\mu}=-\frac{1}{2\pi^2}\epsilon_{\mu\nu\alpha\beta} 
\partial^\mu(A^\nu+\mu u^\nu)\partial^\alpha \mu_5 u^\beta~~.
\end{eqnarray}
Rewriting Eqs. (\ref{anomaly1}) and (\ref{anomaly2})  
\begin{eqnarray}\label{anomcur}
\partial_\mu \left(n_5u^\mu+\frac{\mu^2+\mu_5^2}{2\pi^2}\omega^\mu+\frac{\mu}{2\pi^2}B^\mu\right)=-
\frac{1}{4\pi^2}\epsilon_{\mu\nu\alpha\beta}\partial^\mu A^\nu\partial^\alpha A^\beta\nonumber\\
\partial_\mu \left(nu^\mu+\frac{\mu\mu_5}{\pi^2}\omega^\mu+\frac{\mu_5}{2\pi^2}B^\mu\right)=0~~.
\end{eqnarray}
allows for a straightforward comparison with results of 
the thermodynamic approach, see Eqs.
 (\ref{coeff1}), (\ref{coeff11}). We find out  that the effective theory does reproduce  anomalous pieces
 of the transport coefficients obtained earlier in the leading order in the chemical potentias.
As for the higher orders in the chemical potential
 there are apparent differences.  
Within the effective theory higher orders in the chemical potential
 belong to higher order perturbative terms. 
The triangle graphs which determine (\ref{anomcur})
have of course  a very special status. First, they  defy  conservation of the currents and, second,
they do  not receive contributions due to the iteration of $S_{int}$, because of the
Adler-Bardeen theorem.

 As for  higher in $\mu,\mu_5$ terms they are infrared sensitive 
 and can be fixed only within a particular infrared-sensitive regularization scheme. 
 Consider for example contribution of order $\mu^3$ to the chiral vortical effect.
 It can be estimated as
\begin{equation}\label{estimate}
\delta \xi~\sim~ {\mu^2\over 2\pi^2}{\mu\over \epsilon_{IR}}~,
\end{equation}
where $\epsilon_{IR}$ 
is an infrared cut off in the energy/momentum integration. 
Eq (\ref{estimate}) can hardly be improved within the effective theory.
On the other hand, the thermodynamic derivation of Sect 2.3 does fix 
\cite{surowka}  terms of order $\mu^3$ in the Landau gauge
as:
\begin{equation}\label{twothirds}
\delta \xi~=~-{\mu^2\over 2\pi^2}{2\mu n\over (\epsilon+p)} ~.
\end{equation}
By comparing (\ref{estimate}) and (\ref{twothirds}) we find
\begin{equation}
\epsilon_{IR}~\sim~(\epsilon+p)/n~~
\nonumber~~.
\end{equation}
 Note that the enthalpy $w=\epsilon +p$ is known to play the role
of mass in relativistic hydrodynamics. Thus, the ratio
$\mu\cdot n/(\epsilon+p)$ characterizes the contribution of the energy related to the chemical potential
in units of  the total energy, as far as $\mu$ 
is small. And it is quite natural to have an expansion     
 in this parameter. The expansion coefficients within the effective theory 
are model dependent, however.
 
To summarize, the effective hydrodynamic theory defined through the substitution
(\ref{rule}) allows for a straightforward and simple evaluation of the chiral
effects in terms of anomalies of the effective theory. Terms of lowest order in expansion
in the chemical potentials coincide with the results of other approaches. 
Higher orders, however, are infrared sensitive and model dependent within the effective
theory. Within the thermodynamic approach of Sect. 2.1  these terms are
apparently  fixed by the procedure chosen to integrate the differential
equations (\ref{eqD1})  beginning with small $\mu$ and  keeping the pressure $P$ constant.  
 The use of the Landau frame is also needed.
  
Apart from the $\mu^2$ term the vortical effect has a $T^2$ contribution,
see Sect. 2.3.
This term canbe evaluated within finite-temperature field theory 
at $\mu=0$. For this reason
the effective in chemical potentials theory introduced 
above does not help to evaluate  the $T^2$ term.

\subsection{Concluding remarks}

Powerful non-renormalization theorems 
have been proven in case of chiral magnetic and vortical effects
in hydrodynamic approximation. Eventually, all the proofs go back to
celebrated field theoretic non-renormalization theorems \cite{bardeen,coleman}.
To bridge them to hydrodynamics, it is crucial that
chiral effects can be expressed in terms of  certain 
spatial correlators  at frequency $\omega=0$.
These static correlators are trivially continued to the
Euclidean space. Which means, in turn, that we
are dealing with  thermodynamic observables. 
Continuation to the Euclidean space
also allows to use the standard technique of Feynman graphs and
utilize field-theoretic non-renormalization theorems. 
We also notice that all the theorems refer to topological terms in action.
In other words, the action observes symmetries of the problem 
considered while
the Lagrangian density  does not. Chiral anomalies were put into
such a  context first in Ref. \cite{witten}. Moreover, it turns out that
it is not crucial whether the corresponding 
one-loop graphs signify an anomaly or not. 
Probably, one and the same non-renormalization
theorem can be proven either in terms of anomalies or
non-anomalous graphs. The topological aspect, however, seems to be
an indispensable ingredient to non-renormalizability. 

\section{Hydrodynamic Chiral Effects as Quantum Phenomena}

\subsection{Non-dissipative currents}

All the chiral effects which we are considering are non-dissipative.
This was demonstrated in the thermodynamic language \cite{cheianov,general,jensen},
see Sect. 2.2. Another line of reasoning is suggested  in Ref.   \cite{dissipation} 
and is based on time-reversal invariance.
Let us summarize the argumentation \footnote{The author is thankful to L. Stodolsky
for a detailed discussion of the subject.}. One compares, for example,
ordinary electric conductance, $\sigma_E$ and  $\sigma_B$
associated with the ChME:
$$\vec{ j}_{el}~=~\sigma_E\vec{ E},~~~\vec{ j}_{el}~=~\sigma_B\vec{B}~~.$$
Moreover, $\sigma_E>0$ since the work done by external electric field,
$W~=\vec{j}_{el}\cdot\vec{E}>0$.
 
Under time reversal,
$$\vec{ j}^T_{el}~=~-\vec{j}_{el},~~\vec{E}^T~=~+\vec{E},~~\vec{B}^T~=~-\vec{B}~~.$$
If we would try to obtain $\vec{ j}^T_{el}~=~\sigma_E ^T\vec{E}^T$
by time reversal of the relation $\vec{j}_{el}~=~\sigma_E \vec{E}$
then we would conclude that $\sigma^T_E~=~-\sigma_E$
which is in contradiction with the positivity of $\sigma_E$.
There is no surprise of course in this failure. To the contrary,
dissipation is indeed not a time-reversible process.
(The time-reversal invariance is manifested instead in the Onsager
relations which imply, in particular, $\sigma_E~>~0$).
On the other hand,  the Hall relation $\vec{j}_{el}~\sim~\vec{E}\times \vec{B}$
is time-reversal invariant and this is possible only if there is no dissipation.
 Our chiral magnetic effect is of the same type
as the Hall conductivity and, therefore, cannot be accompanied by dissipation
because of the time-reversal invariance of theories considered.
Note that there are no such symmetry-based arguments for, say, 
superfluid current. The viscosity can be only dynamically suppressed in that case
and according to the modern views \cite{kovtun} there is a universal
lowest bound on the shear viscosity, $\eta/s \ge 1/4\pi$ where
$s$ is the entropy density.   According to the logic outlined above no 
similar bound can exist for dissipation associated with
the chiral magnetic effect since  dissipation is forbidden 
by symmetry considerations.

 Furthermore, dissipation-free processes are usually quantum phenomena
and in this section we discuss the ChME from this point of view. 
In fact, it is rather an open-end discussion 
since not much known yet about the microscopic picture of the 
ChME in the hydrodynamic approximation.   

There is no doubt that the ChME is rooted in the loop graphs of the
underlying theory and is represents a macroscopic manifestation
of quantum phenomena. Calculation of a standard Feynman graph gets related
to thermodynamics by a simple trick
of identifying a constant piece in an external gauge field with
the chemical  potential, $A_0~\approx~\mu$. Let us explain this point in more detail.
 The generating  functional
$W(sources(x))$, see Eq. (\ref{differentiate}), contains in fact
a piece, $W^{(1)}_{anom}(sources(x))$ reproducing the chiral anomaly. 
Moreover, $W_{anom}$ is uniquely fixed by  the  requirement 
that it  does reproduce the anomaly within the formalism
of Sect. 2.2.
It turns out that $W_{anom}$ in the static case
considered  can be written in a local form \cite{general}.
Explicitly to first order in derivatives:
\begin{equation}\label{anomalous}
W^{(1)}_{anom}~=~\frac{C}{2}~\int d^3x \sqrt{g_3}  
\Big({A_0\over 3T_0}\epsilon^{ijk}A_i\partial_jA_k+{A_0^2\over 6T_0}\epsilon^{ijk}A_i\partial_ja_k\Big)~~,
\end{equation}
where the constant $C$ determines the anomaly and further  notations are specified in Sect. 2.2.
 
 Let us emphasize that Eq. (\ref{anomalous}) is a pure field-theoretic input
valid in the  exact chiral limit. However, once we identify $A_0$ with
 a macroscopic  quantity, chemical potential, the action (\ref{anomalous})
 determines macroscopic motions. Moreover, it  is quite obvious that
two terms in the right-hand side of Eq. (\ref{anomalous}) do
 reproduce the chiral magnetic and chiral vortical effects,
  respectively. Comparison of (\ref{anomalous}) with 
 (\ref{sigmab}) and (\ref{xiomega}) helps to check this.

   As is emphasized above, Eq. (\ref{anomalous}) is valid in 
   exact chiral limit. If we would decide to estimate
the effect of chiral symmetry violations, say, through
finite fermionic masses, we have to address
the field-theoretic calculations anew. In particular, 
the rate of production of massive fermions
in parallel constant electric and magnetic fields, $E_z,B_z$ is given by
the equation
\cite{warringa1}:
\begin{equation}
{dN_5\over d^3x dt}~=~{q^2B_zE_z\over 2\pi^2}\exp \big(-{\pi m_f^2\over E_z}\big)~~,
\end{equation}
which is replacing Eq (\ref{tobereplaced}). We see
 that if we tend $E_z\to 0$ now we would not get
 any ChME, compare Eq. (\ref{linear}).
Therefore, the formal static hydrodynamic limit is replaced for small 
fermionic masses by:
\begin{equation}
(\omega~\equiv~ 0, q_i~\to~0)~\rightarrow~(|q_i|~\gg~\omega,~\omega~\gg~m_f)~,
\end{equation} 
and phenomenological implications of  this modification are to be considered 
within a particular framework.

 Quantum description of the chiral magnetic 
 effect has another aspect which can be illustrated on the
 example of chiral fermions interacting with external
 magnetic field \cite{nielsen,warringa}. Namely,  we should be able to obtain
the same
 current (\ref{magnetic})
 by evaluating the matrix element
 \begin{equation}\label{averaging}
 j^{el}_{\mu}~=~q<\bar{\psi}(x)\gamma_{\mu}(x)\psi(x)>~~,
 \end{equation}
 where the averaging is over the thermodynamic ensemble.
 The simplest set up is  non-interacting fermions and a 
non-vanishing chemical potential $\mu_5$.
In this case  $\psi(x)$ represents solutions of the
Dirac equation
\begin{equation}
\big(i\gamma^{\mu}D_{\mu}+\mu_5\gamma^0\gamma^5\big)\psi(x)~=~0~,
\end{equation}
 where $D_{\mu}=\partial_{\mu}-iA_{\mu}$ and the vector-potential
 $A_{\mu}$ corresponds 
 to a constant magnetic field. The thermodynamic ensemble is described 
by the ideal gas of massless Fermi particles.
 Explicit calculation turns to be feasible and the final result agrees
 with (\ref{magnetic}). We will give some details of the calculation
 in Sect. 3.4. Note that, at least from the technical  point of view,
 this coincidence is not trivial at all. Indeed, Eq. (\ref{magnetic})
 is based entirely on the evaluation of the anomalous triangle graph
over perturbative vacuum state.
The magnetic field is treated perturbatively. On the other  
 hand, the matrix element (\ref{averaging}) 
 is saturated by the fermionic zero modes \cite{nielsen} 
 which one finds explicitly,  accounting for the magnetic field
 to all orders.
 
 What is lacking in case of the ChME in hydrodynamic approximation
 is a microscopic calculation similar to the direct evaluation
 of (\ref{averaging}) for non-interacting particles just outlined.
 What makes such a calculation especially desirable is the observation
 that all the non-renormalization theorems reviewed in Sect. 2
 do not indicate any crucial dependence on temperature.  On
 the other hand, well known non-relativistic analogues, like
 the Hall conductivity or ordinary superfluidity do exhibit
 sharp dependence on the temperature. 

In the rest of this section we discuss microscopical picture
for the chiral effects on the only example available so far \cite{kirilin},
that is the case of superfluid.

\subsection{Low-dimensional defects}

A novel point brought by considering superfluid is the 
crucial role
of low-dimensional defects, or singularities of hydrodynamic approximation.
Such defects were considered in fact in many papers
for various reasons, see in particular
\cite{goldstonewilczec,callanharvey,zhitnitsky,witten10}. In case
of rotating superfluid such defects have been known since long \cite{landau9}.

Consider first the chiral vortical effect.
The crucial point is that
 the velocity field of the superfluid is known to be potential:
 \begin{equation}\label{varphi}
 \vec{v}_s~=~\vec{\nabla}\varphi~,
 \end{equation}
 and, 
naively, the vorticity vanishes identically
since $curl \vec{v}_s=0$. If this were true, the chiral current
 (\ref{vortaic}) would disappear.  But it is well known, of  course, that the angular
momentum is still transferred to the liquid through vortices \cite{landau9}.
The potential is singular on the linear defects, or  vortices.
The vortex is defined through circulation of velocity:
\begin{equation}\label{quantization}
\oint~\vec{v}_sd\vec{l}~=~2\pi k~,
\end{equation}
where $d\vec{l}$ is an element of length
and $k$ is integer. The quantization condition (\ref{quantization}) follows from
the interpretation of $\varphi$ as the phase of a wave function of identical
particles. Eqs. (\ref{quantization}), (\ref{varphi}) imply that the velocity is singular
at the "beginning of the coordinates". In three dimensions 
the singularity occupies a line, which can be
called a defect of lower dimension,
or vortex. The energy of the vortex is logarithmically divergent,
$E_{vortex}~\sim~l\ln (l/a)$ where $l$ is the length of a (closed) vortex
and $a$ is  of order distance between the constituents.
 
Note that the velocity $v_s$ according to (\ref{quantization}) falls
off as a function of distance $r$ to  the singularity,  $v_s~=~k/r$
while for a rotating solid body the velocity, to the contrary, would grow
with  distance to the axis of  rotation, $v= |\vec{\Omega}|r$
where $\vec{\Omega}$ is the angular velocity of rotation. Therefore,
at first sight, distribution of  velocities inside a rotating 
bucket with a superfluid is very different from  ordinary
liquid. This is not true, however, as far as the angular velocity
 is large enough. In this case, there are many defects and the
 distribution of velocities, once averaged over defects, is the same as for ordinary
liquid. The proof \cite{landau9} is based on the observation
that thermodynamic equilibrium  is determined by minimizing
$E_{rotation}~=~E-\vec{M}\cdot\vec{\Omega}$ where $\vec{ M}$ is angular
momentum and the averaged (over defects) velocity is uniquely
determined by this condition. 
 
 To summarize, vortices in superfluid represent a well-known example
 of lower-dimensional defects. Although locally, or at small distances,
 defects look  very different from the hydrodynamic 
 picture, the thermodynamic results can be restored upon averaging over a large number
 of defects.

\subsection{Relativistic superfluidity}
 
 To consider vortex solutions in detail  we need explicit
 examples of dynamical systems which exhibit relativistic superfluidity.
 The simplest and  best understood example
\cite{sonstephanov}  of this type  seems to be the pion medium 
at zero temperature  and non-zero isospin chemical potential $\mu_I$.
The system is described by  the chiral Lagrangian;
\begin{eqnarray}
\label{lag}
L=\frac{1}{4}f_\pi^2Tr[D^{\mu}U(D_{\mu}U)^\dag], 
\end{eqnarray}
where $U$ are $2\times 2$ unitary matrices, functions of the pionic fields, see, e.g.,
\cite{leutwyler}.
Moreover,
the chemical potential $\mu_I$ is switched on through the covariant derivatives,
$D_0U=\partial_0 U-\frac{\mu_I}{2}[\tau_3, U]$, $D_iU=\partial_iU$. 
To have the chiral current conserved we consider massless quarks. Then
the chiral symmetry is spontaneously broken. In the common case of vanishing
chemical potential the residual  symmetry (realized linearly)
 is $SU(2)_{L+R}$. Furthermore, at non-zero $\mu_I$ this symmetry 
is broken to $U(1)_{L+R}$. The proof is straightforward.
Namely,
the potential energy corresponding to (\ref{lag}) equals to
\begin{eqnarray}
V_{eff}(U)=\frac{f_\pi^2\mu_I^2}{8}Tr[\tau_3U\tau_3U^\dag-1],
\end{eqnarray}
and the 
minima of that potential can be captured 
by substitution $U=\cos{\alpha}+i(\tau_1\cos{\phi}+\tau_2\sin{\phi})\sin{\alpha}$:
\begin{eqnarray}
V_{eff}(\alpha)=\frac{f_\pi^2\mu_I^2}{4}(\cos{2\alpha}-1)
\end{eqnarray}
and for the minimum one readily obtains $\cos{\alpha}=0$. Then, depending on the sign of
 $\mu_I$, squared mass of $\pi^+$ or $\pi^-$ state becomes 
negative and the corresponding field 
is condensed. This means that the vacuum is described by 
$U = i(\tau_1\cos{\phi}+\tau_2\sin{\phi})$ instead of the standard, i.e. $\mu=0$   
vacuum $U = I$.
There emerges 
a new order parameter
 $\langle \overline{u} \gamma_5 d\rangle + h.c. = 
2 \langle\overline{\psi}\psi\rangle_{vac} \sin(\alpha) = 2 
\langle \overline{\psi}\psi\rangle_{vac}$. 
The system is thus a charged superfluid. It should be noted, that the 
degeneracy with respect to the angle $\phi$ above
indicates that it can be identified as a 3d Goldstone field.
In addition, there are two massive modes.

Because of the presence of a 3d Goldstone mode
 the superfluidity criterion (\ref{3d}) is satisfied:
\begin{equation}\label{criterion}
\lim_{\vec{q}\to 0}{\int d^3x~\exp(i\vec{q\cdot r})~<T_{0i}(\vec{x}), T_{0k}({0})>}
~=~\mu^2\vec{ q}_i\vec{q}_k/\vec{ q}^2 ~.
\end{equation}
Explicit evaluation of this correlator
is based on the Josephson equation:
\begin{equation}\label{josephson}
\partial_0\phi~=~\mu~
\end{equation}
which is satisfied now as an equation of motion following from (\ref{lag}). For 
$\mu=const$
  Eq. (\ref{josephson}) can be interpreted as  condensation
of $\partial_0\phi$, similar to the standard Higgs condensation but violating the
Lorentz invariance \cite{son1,nicolis1}.  


  As is argued, e.g., in 
\cite{son1}, $\partial_\mu\phi$ can be identified with non-normalized superfluid velocity.
The vortex configuration is in principle determined by the value of the angular velocity
of rotation of the liquid
\cite{landauV}.
 We address a generic  case, when the quantum of circulation, $n$ 
is rather high to consider defects thermodynamically
 but not high enough to ruin the superfluidity. 
As is known, an energetically preferable configuration is the uniform distribution 
of vortices with $n = 1$. 
Nearby any given vortex the Goldstone field is given by \cite{nicolis1}:

\begin{eqnarray}
\phi=\mu t+\varphi,
\end{eqnarray}

We will assume that vortices are well separated, $\delta x >> a$, calculate 
the current for a single vortex and then sum it over all vortices, that is 
simply multiply by $n$.
The effective Lagrangian for the interaction of fermions 
with the scalar field $\phi$ (we will limit ourselves  to the case of single fermion, and then sum up  
the  result over colors and flavors) looks as
\begin{eqnarray}
L=\overline{\psi}i(\partial_\mu+i \partial_\mu\phi)\gamma^\mu\psi.
\end{eqnarray}
Indeed because, of the Josephson equation (\ref{josephson}) we reproduce
the standard chemical potential term. Other components complete this term
to a formally Lorentz invariant interaction,  compare Eq. (\ref{rule}).
  
Using standard methods of evaluating the anomalous triangle diagrams,
see, e.g., Sect. 2.4, one obtains 
for the axial current:  
\begin{eqnarray}
j^5_\mu=\frac{1}{4\pi^2}\epsilon_{\mu\nu\alpha\beta}\partial^{\nu}\phi \partial^{\alpha}\partial^{\beta}\phi ,
\end{eqnarray}
This current seems to vanish identically. However,  for the vortex field, 
$\phi = \mu t + \varphi$, $[\partial_x,\partial_y]\phi=2\pi \delta(x,y)$  and:
\begin{eqnarray}
(j^5_3)_{vortex}~ =~ \frac{\mu}{2\pi}\delta(x,y)~~.
\end{eqnarray}
The total current, or
the sum over the vortices equals to
\begin{equation}\label{j}
j^5_3 = \int d^2x j^5_3 = \frac{\mu}{2\pi}n~.
\end{equation}
It is worth noting that actually $n~\sim~\mu$ and the current (\ref{j})
is quadratic in the chemical potential $\mu$, as it should be.

So far we considered the chiral vortical effect.
 To evaluate the chiral magnetic effect, consider a 
 charged superfluid and turn on a magnetic field. 
Then magnetic field would stream into tubes, or Abrikosov vortices.
The vortex profile  could be found by accounting for the finite
photon mass.
The chiral
current can  be obtained then
 by substituting the vortex configuration to the
Dirac equation and solving it for the modes. The current is concentrated
 on the vortex
center.  In the hydrodynamic approximation we are considering
the magnetic field is singular and the Dirac equation is poorly defined in this sense.
However, using index theorems it is possible to evaluate number of zero
modes, and the zero modes saturate the chiral current.
 We will give more details in the next subsection.

\subsection{Zero modes}
 We now proceed  to  microscopic picture based on the zero modes
 \footnote{This subsection is of rather technical nature and can be considered
 as an appendix.  Moreover, the presentation is close to that of Ref.
  \cite{zhitnitsky},
 with a substitution    $A_{i} \rightarrow \partial_i \phi$. } .  
The Hamiltonian has the form :
\begin{eqnarray}
H = -i (\partial_i - i  \partial_i \phi)\gamma^0 \gamma^i + m\gamma^0
\end{eqnarray}
and the Dirac equation decomposes as:
\begin{eqnarray}
-H_R\psi_L = E \psi_L\\
H_R \psi_R = E \psi_R,
\end{eqnarray}
where $H_R = (-i\partial_{i} + \partial_{i} \phi) \sigma_{i}$.
Hence, any solution $\psi_R$ of $H_R \psi_R = \epsilon \psi_R$ 
generates both  a solution with energy $E = \epsilon$,
$
\psi=\left(\begin{array}c
0\\
\psi_R
\end{array}\right)
$ and a solution with $E = -\epsilon$,
$
\psi=\left(\begin{array}c
\psi_{R}\\
0
\end{array}\right)
$.

Because of the invariance with the respect to translations in z direction, 
we decompose using the 
momentum eigenstates $-i \partial_3 \psi_R = p_3\psi_R$.   For each $p_3$,
\begin{eqnarray}\label{equations}
H_R &=& p_3 \sigma^3 + H_\perp \\
H_\perp &=& (-i \partial_a - \partial_a \phi) \sigma^a, \;\; a = 1,2.
\end{eqnarray}
Notice that $\{\sigma^3, H_\perp\} = 0$.  
This means that if $|\lambda\rangle$ is an eigenstate of $H_\perp$ with non-zero 
eigenvalue  $\lambda$, $H_\perp |\lambda\rangle = \lambda |\lambda\rangle$ ,
$\sigma_3 \lambda\rangle$ is an eigenstate with eigenvalue $-\lambda$.
are of the form 
$|\lambda\rangle$, 
$|-\lambda\rangle = \sigma_3 |\lambda\rangle$, with $\lambda > 0$. Also, $\sigma_3$ 
maps zero eigenstates of $H_\perp$ to themselves, 
so that all the eigenstates of $H_\perp$ can be 
classified with respect to $\sigma_3$.

 We can now express eigenstates of $H_R$ in terms of eigenstates of $H_\perp$. 
Since $[H_R,{H_\perp}^2] = 0$, $H_R$ will only mix states $|\lambda\rangle$, 
$|-\lambda\rangle$. For $\lambda > 0$, one can write,
 \begin{eqnarray}
 \psi_R = c_1|\lambda\rangle + c_2 \sigma_3|\lambda\rangle.
 \end{eqnarray}
 Solving equations (\ref{equations}) we find for the eigenvalues of energy:
  $$\epsilon = \pm \sqrt{\lambda^2 + p_3^2}~.$$  
 This means that every eigenstate of $H_\perp$ with eigenvalue $\lambda > 0$ 
produces two eigenstates of $H_R$, while  zero modes of $H_\perp$,
$|\lambda=0\rangle$  are
 eigenstates of $H_R$ with eigenvalues:
 \begin{eqnarray}
 \epsilon~ = ~\pm p_3~,
 \end{eqnarray}
where  the sign corresponds to $\sigma^3|\lambda=0\rangle~=~\pm |\lambda=0\rangle$.

 Therefore, the zero modes of $H_\perp$ are gapless modes of $H$, capable of traveling  
up or down the vortex, depending on the sign of $\sigma_3$ and
 chirality. These are precisely the carriers of the axial current along the vortex.  
Let $N_{\pm}$+ be the numbers of zero modes
which are eigenstates of the matrix $\sigma^3$ with eigenvalues $\pm 1$,
 respectively. Consider  zero mode of $H_\perp$, 
$|\lambda\rangle = (u,~ v)$, where $u$ and $v$ are c-functions satisfying
 \begin{eqnarray}
 {\cal D} v = 0, ~~~ {\cal D}^{\dag} u = 0.
 \end{eqnarray}
 Here
 \begin{eqnarray}
 {\cal D} = -i \partial_1 - \partial_2  -  (\partial_1 \phi - i \partial_2 \phi).
 \end{eqnarray}
Define
  $N_+ = dim(ker({\cal D}^{\dag}))$, $N_- = dim(ker({\cal D}))$, 
and
 \begin{eqnarray}
N = Index(H_\perp) = N_{+} - N_{-} =  dim(ker({\cal D}^{\dagger})) - dim(ker({\cal D}))
 \end{eqnarray}

Note that $H_\perp$ is an elliptic operator. Its index has been computed within
various approaches in papers \cite{aharonov}. In our case the index is given by
\begin{eqnarray}\label{result}
N = \frac{1}{2\pi}\int d x_{i} \partial_i \phi = n~.
\end{eqnarray}
Moreover, for the most important case $n=1$ the zero mode is easy to construct,
see, e.g., Ref. \cite{kirilin}.
The result (\ref{result}) can be also obtained starting from 
the well known case of magnetic field
parallel to z-axis and uniform in that direction. In the latter
case the index is given by  
\begin{eqnarray}\label{result1}
N = \frac{e}{2\pi}\int d x_{i} A_{i} = \frac{e}{2\pi} \int d^2 x  B_z
\end{eqnarray}
and by  substituting
$e A_{i} \rightarrow \partial_{i} \phi$ we arrive at (\ref{result}).
Note, however,  that in  case of superfluid, which we discuss here, the index 
an integer, 
whereas for non-superconducting case the flux  is not quantized
and the left-hand side of Eq. (\ref{result1}) is to be understood as the integer
part of the right-hand side.

We now proceed to the computation of the fermion axial current at a
 finite chemical potential $\mu$. The axial current density in the third direction
is given by:
\begin{eqnarray}
j^3_5(x) = \overline{\psi}(x)\gamma^3 \gamma^5 \psi(x) = 
\psi_L^{\dag} \sigma^3 \psi_L(x) +\psi_R^{\dag} \sigma^3 \psi_R(x)
\end{eqnarray}
We are interested in the expectation value of the
 axial current along the vortex, 
$j^3_5 =\int d^2x \langle j^3_5(x)\rangle$. At finite chemical potential, we have:
\begin{eqnarray}
\langle j^3_5(x)\rangle = 
\sum_{E} \theta (\mu - E)\,\psi^{\dagger}_E(x) \gamma^0 \gamma^3 \gamma^5 \psi_E(x)= \\ \nonumber
\sum_{\epsilon} (\theta(\mu - \epsilon) + 
\theta(\mu + \epsilon)))\psi_{R\epsilon}^{\dagger}(x)\sigma^3 \psi_{R\epsilon}(x)
\end{eqnarray}
Here, $\theta (\mu - E)$ is the Fermi-Dirac distribution (at zero temperature), 
$\psi_E$ are eigenstates of $H$ with eigenvalue $E$, $\psi_{R \epsilon}$ are eigenstates of 
$H_R$ with eigenvalue $\epsilon$. By substitution of the explicit form of 
$\psi_{R\epsilon}$ in terms of $H_\perp$ eigenstates, one obtains:
\begin{eqnarray}
\langle j^3_5\rangle~=~
\frac{1}{L}\sum_{p_3}\sum_{\lambda
>0}\sum_{s = \pm}(\theta(\mu - (\lambda^2 + p_3^2)^{\frac{1}{2}}) +\\ \nonumber
\theta(\mu + (\lambda^2 + p_3^2)^{\frac{1}{2}}))
\langle\psi^s_R(\lambda,p_3)|\sigma^3|\psi^s_R(\lambda,p_3)\rangle
+
\\ \nonumber \frac{1}{L}\sum_{p_3}\sum_{\lambda = 0} (\theta(\mu - p_3)+\theta(\mu + p_3)) \langle
\lambda |\sigma^3|\lambda\rangle.
\end{eqnarray}

 Here $\lambda > 0$ enumerate eigenstates of $H_\perp$, which generate eigenstates of $H_R$, 
$\psi^\pm_R(\lambda,p_3)$ with momentum $p_3$
and eigenvalue $\epsilon_{\pm} = \pm \sqrt{\lambda^2 + p_3^2}$, and $\lambda = 0$ 
label the zero modes of $H_\perp$. Moreover, the sum over all non-zero eigenstates vanishes, 
and only zero modes of $H_\perp$ generate $j^3_5\neq 0$. For the zero modes, 
$\langle \lambda|\sigma^3|\lambda\rangle = \pm 1$, and we obtain:
\begin{eqnarray}
j^3_5 =
(N_+ - N_-) \frac{1}{L}\sum_{p_3}(\theta(\mu - p_3) +
\theta(\mu + p_3))= \\ \nonumber
= n  \int \frac{dp_3}{2\pi} (\theta(\mu - p_3) + \theta(\mu+ p_3)) = \frac{\mu}{\pi} n
\end{eqnarray}
 
This result is similar to the macroscopic answer (\ref{vortaic}) for the
vortical effect  but there is inconsistency of a factor of two.   

\subsection{Concluding remarks}

Considering superfluid provides a unique  possibility
to develop a   microscopic picture
for chiral hydrodynamic effects.
The calculations above demonstrate that the chiral currents are
carried by quantum-mechanical zero modes
and are indeed dissipation-free.
 This result is in agreement with the expectations.

However,  this explicit example brings also  new lessons.
First of all, the  prediction 
for the chiral vortical  effect is changed by  a factor  of two
and it is  instructive to appreciate the reason for  this  change.
Technically, the easiest way to trace the origin of this factor of two
is to compare
the calculation of the chiral vortical effect in this section 
with the calculation within effective theories, see Sect. 2.4.  In the latter case,
the chiral effects are described by anomalous triangle graphs,
with vertices proportional to $\mu u_{\mu}$ or $qA_{\mu}$.
In other words, the chemical potential $\mu$ plays the role
similar to
the electromagnetic coupling, or charge $q$ while the field of fluid velocities,
$u_{\mu}$ is similar to the electromagnetic field $A_{\mu}$.
The triangle graphs for the chiral magnetic and vortical effects looks
  very similar. The only  difference is that vortical effect is  quadratic
  in $\mu$ while the magnetic effect is linear both in $q$ and $\mu$.
  Because of  quantum mechanics, however, this difference
  results in a factor  of two: in case of the vortical effect
  the corresponding  graph has two identical vertices and
  this brings a factor  of 1/2, as usual. It is this  factor
  which is  absent from the calculation of the  vortical effect
  in terms of defects.
  Indeed, there are two  facets of  the chemical potential.
  First, it plays the role of an effective coupling, as we have just explained.
  And, second, it limits the integration over the longitudinal momentum
  of zero modes, see Sect. 3.4. In the language of defects, these two
   roles are  not interchangeable and the quantum-mechanical factor of
   1/2 of the effective theory is  not reproduced.
   
   Capitalizing on this technical explanation,
   we can say that in terms of  defects 
   we have a two-component picture.  One component
   is superfluid with velocity field $u_{\mu}$.
   The other component are zero modes responsible for the
   chiral magnetic and vortical effects. The zero
   modes are having speed of light and, therefore,
   are not equilibrated with the rest of the liquid. 
The two-component
picture, however, does not necessarily differ in 
predictions from one-
component picture.
Indeed, we did reproduce the standard answer in the
case of chiral  magnetic effect. Technically, the
reason is the same as in Sect. 4.1 where
we argued (following Ref. \cite{landau9})
that vortices reproduce on average the velocity
distribution of  ordinary rotating liquid.
Namely, linearity of the ChME effect in the chemical
potential is the same crucial as linearity in the
angular momentum of the energy $E_{rot}$ in case
of rotating liquid.

\section{Conclusions}
 
  Theory of the chiral  magnetic effect has been developing fast 
  since the papers \cite{warringa} put it  into the actual context
  of the RHIC experiments. At the beginning,
theory was focused on the mechanism  of chirality
production in heavy ion collisions. Already at this stage
one has to turn to hydrodynamics since it 
  describes the bulk  of  the  RHIC data.   
 Since effective chiral chemical potential $\mu_5$ 
vanishes on average and 
 fluctuates from event to event in heavy ion collisions,
 it is mostly physics of fluctuations 
 which--in theoretical perspective-- was studied at this stage.

Later, beginning with the paper \cite{surowka} the interest    
  shifted from phenomenology to more theoretical  issues,
  such as unifying methods of field theory and thermodynamics to get
  exact results for  chiral  effects in hydrodynamic approximation.
  Very recently, we believe, the  most exciting development 
  is the emerging proof that chiral  magnetic effect in hydrodynamics
  is a  dissipation-free process. Moreover, examples known so far indicate that
this ballistic-type of transport is provided by 
quantum-mechanical zero modes.

All these exciting results are valid, strictly speaking, in the exact chiral limit.
 This limitation might have rather severe phenomenological implications
in case of realistic quantum chromodynamics. One could expect, therefore,
in the near future a shift of interest to condensed-matter systems
with fermionic excitations and linear dependence of the energy on
 the momentum. Of course, this classification of theoretical developments
into various stages   at the very best could be true only in its gross features.
Nevertheless, it might be reasonable to present our conclusions within this,
oversimplifying
scheme.  

{\bf Fluctuations of the chiral chemical potential}. Because of the space
limitations we did not address the issue of fluctuations in the bulk of the review.
If we start with $<\mu_5>=0$
the current (\ref{magnetic}) vanishes.
The chiral magnetic effect is still manifested
through fluctuations.  
In particular,  the spectrum of hydrodynamic excitations
is sensitive to it. 
 The so called chiral magnetic wave \cite{kharzeev11}
 corresponds
 to the following sequence of fluctuations:
 \begin{equation}
 (\delta n_Q)~\to~(\delta j^5)~\to~(\delta \mu_A)~\to~(\delta j^{el})~\to~(\delta n_Q)~~.
  \end{equation} 
In more detail: first, a local fluctuation of electric charge density  
induces fluctuation of axial current, see (\ref{vector}). Then the fluctuation
of the axial current triggers a local fluctuation of the
axial chemical  potential. Finally and completing the cycle, the fluctuation
of $\mu_5$ results in a fluctuation of the electrical current, see (\ref{magnetic}).
Thus, there should exist an excitation combining density waves of 
electric and chiral charges, the chiral magnetic wave.

Another result to be mentioned here is the observation
\cite{shevchenko1} that there is a piece in the correlator
of components of electric current which is uniquely determined in
terms of the chiral anomaly squared:
\begin{equation}
F_{zz}(\omega)~-~F_{xx}(\omega)~=~
\frac{(qB)^2}{4\pi^3}\frac{\omega}{e^{\beta \omega}-1}~~,
\end{equation}
where the $z$-axis points in the direction of an external magnetic field
and $F_{ii}(\omega)$ is  a correlator of the $i$-th components of the
electromagnetic current as a function of frequency $\omega$, for a precise definition 
of the correlators see \cite{shevchenko1}.

{\bf Non-renormalization theorems and non-dissipative motions.}
The present review is focused on the non-renormalization theorems
and non-dissipative, or  quantum  nature of the chiral effects,
 see Sect. 2 and Sect. 3, respectively. Non-renormaliization theorems 
   were proven within various approaches (thermodynamic,  geometric,
   diagrammatic, effective field theories).  There are no doubts that the
      non-renormalization theorems are valid within the approximations and assumptions made.
      The basic assumptions are exact chiral limit,
hydrodynamic  expansion in derivatives 
       and equilibrium. The main message  is that chiral currents are dissipation
free and there is  no suppression at high temperature.  

The results reviewed in Sect. 2 imply that the dissipation-free 
motions considered are rather ballistic transports than superfluid-type
phenomena. Indeed, the entropy current associated with the chiral effects
is not vanishing, unlike the superfluid case. According to (\ref{eqD2}), (\ref{coeff11})
$$s_{\mu}~\sim~\frac{\mu}{T}\mu_5B_{\mu}~~,$$
where $s_{\mu}$ is the entropy current in equilibrium associated
with the magnetic field $B_{\mu}$ and $\mu$, $\mu_5$ are chemical potentials.
The entropy current disappears for $\mu=0$ but this is actually a kind of misleading.
The point is that there is always a flow of degrees of freedom along
 the magnetic field and in the direction opposite to it. There is a cancellation,
in terms of $s_{\mu}$
at $\mu=0$ while $\mu\neq 0$ implies that the liquid is charged
and there is preferred direction of the flow of degrees
of freedom. The electric current (\ref{magnetic}), on the other hand, counts the total number
of degrees of freedom with charge $\pm q$.    

Moreover, at least in case of noninteracting particles the carriers of
the electromagnetic current (\ref{magnetic})
are identified as quantum-mechanical zero modes $\psi_0$
of Dirac particles in the external magnetic field $B$: :
$$D_{\mu}\gamma^{\mu}\psi_0~=~0,~~N_0~\sim~|q\cdot B\cdot \mu_5|~,$$
where $N_0$ is the number of zero modes, see Sect. 3.4 or, e.g., 
\cite{dunne} and references therein.
Theoretically, no suppression of the current (\ref{magnetic}) is expected
at non-zero temperatures. A straightforward conclusion would be that the number
of zero modes does not go down with temperature. At first sight, it looks very
unexpected that quantum coherence could persist at high temperature.  
Let us mention, however, that something similar happens already at $T=0$.
Namely, measurements at small lattice spacings $a$ (such that $a\cdot \Lambda_{QCD}\ll 1$)
demonstrate that the number of zero modes survive wild quantum fluctuations
which are of order $|A^{glue}_{\mu}|~\sim~1/a$. However, the volume $V_0$ occupied by
a zero mode goes to zero as a power of $(a\cdot \Lambda_{QCD})$,
for a review see, e.g., \cite{viz}. By analogy, one could expect
for zero modes at high temperature
$$N_0(T)~\approx~const,~~~V_0~\approx~(\Lambda_{QCD}/T)^{\gamma}~~,$$
where the index $\gamma~\sim~(1-2)$, and zero modes at high temperature
become a kind of lower-dimensional defects discussed in Sect. 3.3.

Technically, derivation of dissipation-free hydrodynamics from
chiral symmetry of field theory is  quite straightforward. 
Consider first confining phase. In field theory,
there is spontaneous symmetry breaking and light degrees of freedom are represented
by massless Goldstone fields $\varphi$. This is  field theoretic input. The 
hydrodynamic output
is superfluidity \cite{sonstephanov}  associated  with an
extra
thermodynamic potential (superfluid 3d velocity squared). The route from field theory
to hydrodynamics is provided by replacing ordinary time derivative with a covariant one,
the chemical potential being a constant part of the gauge field $A_0$ As a result,
4d massless Goldstone particle on the field-theoretic
side   becomes a 3d massless  fields (plus the Josephson condition, $\partial_0\phi=\mu$).
Gradient of $\varphi$, $\nabla\varphi=\vec{v}_s$ 
is identified with (unrenormalized) superfluid velocity, or a new thermodynamic variable.
Original fermionic degrees of freedom are counted as constituents of the normal component.
Standard relativistic superfluidity is reproduced.  

In this review, we are concerned with the  case when there 
is no spontaneous breaking of the chiral  
       symmetry. Still, there are massless particles, the chiral  fermions themselves.
       Field  theoretic input is  then existence of polynomials in the effective action
       such that the action observes the  symmetries 
while the density of the action does not
       respect the symmetries.     
       The bridge to hydrodynamics is provided by the same inclusion 
of the chemical potential
       into the covariant derivative. As a result,  there are new motions,
       or currents allowed in the equilibrium
       (similar to superfluidity).  However, because we do not introduce 
          new massless degrees of freedom, the dissipation-free motions are 
calculable now, and there are no new thermodynamic potentials.
(For attempts to introduce chiral supefluidity in direct
analogy with the ordinary superfluidity, i.e. through postulating a new light scalar
see Refs \cite{kalaydzhyan}).

In this sense, analogy between the ChME and Casimir forces, 
(for review and references see, e.g., \cite{casimir}) seems to
be more relevant. One considers first interaction of two fluctuating dipoles  
(atoms)
with electric  polarizabilities
$\vec{p}_{1,2}(\omega)~=~\alpha_{1,2}(\omega)\vec{E}(\omega)$.
At large distances retardation becomes important and
 the van der Waals potential is replaced by the Casimir-Polder
static energy:    
$$V_{CP}~=~-\frac{23}{4\pi}\frac{\alpha_1(0)\alpha_2(0)}{r^6}\frac{hc}{r}$$ 
Note similarity with evaluation of the ChME: in both cases 
static quantities are considered. However, accounting for time dependence
in intermediate state is crucial. The Casimir-Polder potential
refers to interaction of point-like  sources
due to two-photon exchange. Macroscopic interaction
arises upon averaging over many point-like sources. Probably
the best known example of this type  is the force $F$ 
acting on a unit area $A$ of two
conducting plates at distance $a$;
$$\frac{F}{A}~=~-\frac{\pi^2hc}{240a^4}~~.$$
This is a macroscopic force of quantum origin, calculable 
from first principles. Similarly, the anomaly derived first
for point-like particles becomes a macroscopic chiral magnetic effect
 upon
averaging of a two-fermion exchange  over many centers.

{\bf Towards condensed-matter applications.}
A specific feature of the ChME is that even at equilibrium
there is a non-vanishing current while Casimir forces are static.
This  difference is entirely due to the fact that chiral particles
are massless and cannot be  "stopped". Indeed the Casimir-Polder potential
above is due to polarizabilities.
As far as a constituent is massive we can have static magnetic dipole
$\vec{m}~=~\alpha_M\vec{H}$. However, for a massless, chiral  particle
there can be no  static magnetic moment and, instead, we get a
current, see (\ref{magnetic}). 

Already from this simple  reasoning we can conclude that  transition
to chiral effects is singular. If a small mass $m_f\neq 0$ is introduced
the  life-time $T$ of the current (\ref{magnetic}) is finite
and there is no motion in equilibrium, $t\to \infty$, see Sect. 3.1.  Whether
$$T~\sim~\frac{1}{m_f}~~or~~ T~\sim~m_f^2/\mu_5~~,$$
or else,
 remains, to our knowledge, an open question.
The answer might depend on details of experiment
  
It seems natural  to assume that the proof of the non-dissipative nature of
  the chiral effects would be generalized to the case of condensed  
  matter systems, like, say, Weyl semimetals with chiral spectrum of excitations
  $$\epsilon_f~=~vk_f~~,$$
where $\epsilon_f,k_f$ are energy and momentum of fermionic excitation.  
   Indeed, many consequences from the chiral anomaly
   in relativistic field theory have close parallels in  condensed-matter
   systems, see, in particular \cite{condensedmatter}. From the point of view 
   of applications the condensed-matter systems seem of course more practical
and one can start discussing principles of functioning of new kind of  devices \cite{electronics}.
     
       \subsection*{Acknowledgments}
       The author is grateful to T. Kalaydzhyan, D.E. Kharzeev, V.P Kirilin, A.V. Sadofyev, V.I. Shevchenko,
L. Stodolsky  and H. Verschelde  for illuminating discussions.
The work of the author was partially
      supported by grants PICS-12-02-91052, 
FEBR-11-02-01227-a and by  the  Federal Special-Purpose Program
      "Cadres" of the Russian Ministry of Science and Education.

\end{document}